\documentclass[a4paper,11pt]{article}

\usepackage[utf8]{inputenc}
\usepackage[T1]{fontenc}
\usepackage[english]{babel}
\usepackage[top=3cm,bottom=3cm,left=2cm,right=2cm]{geometry}
\usepackage{setspace}
\usepackage{indentfirst}
\usepackage[nottoc]{tocbibind}
\usepackage[affil-it]{authblk}
\usepackage[pdftex]{hyperref}
\usepackage{xcolor}
\usepackage{cite}
\usepackage{amsmath}
\usepackage{amsfonts}
\usepackage{amssymb}
\usepackage{amsthm}
\usepackage{mathtools}
\usepackage{enumitem}
\usepackage{graphicx}
\usepackage{tikz}
\usepackage{orcidlink}
\usepackage{subcaption}
\usepackage{adjustbox}
\usepackage{booktabs}
\usepackage{tabularx}

\definecolor{darkgreen}{rgb}{0,0.7,0}
\definecolor{mahogany}{RGB}{192,64,0}
            
\hypersetup{pdftitle={Theoretical and Experimental Constraints on Z2n Multi-Component Dark Matter Models},
            pdfauthor={J. P. Carvalho-Corrêa, I. M. Pereira, B. L. Sánchez-Vega, A. C. D. Viglioni},
            pdfsubject={High Energy Physics - Phenomenology},
            pdfkeywords={Beyond the Standard Model, Direct Detection, Multi-Component Dark Matter, Perturbative Unitarity, Relic Density, Renormalization Group Equations, Vacuum Stability, Z2n Symmetry},
            pdfcreator={pdfLaTeX},
            pdfproducer={LaTeX with hyperref},
            bookmarksnumbered,
            colorlinks=true,linkcolor=red,citecolor=darkgreen,urlcolor=blue}

\usetikzlibrary{matrix,shapes.geometric,decorations.pathmorphing,decorations.markings}

\numberwithin{equation}{section}

\makeatletter
\def\@maketitle{%
  \newpage
  \begin{center}%
  \let \footnote \thanks
    {\LARGE \@title \par}%
    \vskip 1.5em%
    {\large
      \lineskip .5em%
      \begin{tabular}[t]{c}%
        \@author
      \end{tabular}\par}%
    \vskip 1em%
    {\large \@date}%
  \end{center}%
  \par
  \vskip 1.5em}
\makeatother

\title{Theoretical and Experimental Constraints on $\mathbb{Z}_{2n}$ Multi-Component Dark Matter Models}

\author{J. P. Carvalho-Corrêa\thanks{Email: \href{mailto:jpcarv-15897@ufmg.br}{jpcarv-15897@ufmg.br}} \orcidlink{0009-0009-4420-771X}}
\author{I. M. Pereira\thanks{Email: \href{mailto:isaacmp@ufmg.br}{isaacmp@ufmg.br}} \orcidlink{0000-0002-7203-6737}}
\author{B. L. Sánchez-Vega\thanks{Email: \href{mailto:bruce@fisica.ufmg.br}{bruce@fisica.ufmg.br}} \orcidlink{0000-0002-6735-5813}}
\author{A. C. D. Viglioni\thanks{Email: \href{mailto:arthurcesar@ufmg.br}{arthurcesar@ufmg.br}} \orcidlink{0009-0005-1684-6419}}

\affil{Departamento de F\'isica, UFMG, Belo Horizonte, MG 31270-901, Brazil}

\date{\vspace{-1.615cm}}

\begin{document}

\maketitle

\begin{abstract}
\addcontentsline{toc}{section}{Abstract}

A complete assessment of any dark matter model requires confronting its low-energy phenomenology with its high-scale theoretical viability. We undertake such a dual analysis for a class of two-component scalar dark matter models stabilized by $\mathbb{Z}_{2n}$ symmetries, specifically the $\mathbb{Z}_4$, $\mathbb{Z}_6(23)$, and $\mathbb{Z}_6(13)$ frameworks. Each model is tested against the latest observational data, including the Planck relic abundance and stringent direct detection limits from the LUX-ZEPLIN (LZ) experiment. Simultaneously, we evaluate their theoretical integrity up to the GUT and Planck scales by enforcing vacuum stability and perturbative unitarity with one-loop Renormalization Group Equations. This combined approach reveals a rich and varied landscape of possibilities. We demonstrate that the $\mathbb{Z}_4$ model offers a broadly viable parameter space sustained by efficient semi-annihilation. In stark contrast, the $\mathbb{Z}_6(13)$ scenario is shown to be highly fine-tuned, with solutions confined to the Higgs resonance. Our most significant finding concerns the $\mathbb{Z}_6(23)$ model: we show that an apparent conflict between experimental data and high-scale consistency is resolved when the model is viewed as an effective field theory, yielding a concrete prediction for new physics at or below the $10^6$ GeV scale. This work provides a definitive guide to the viability of these $\mathbb{Z}_{2n}$ scenarios and serves as a compelling demonstration of how high-energy consistency checks can yield crucial insights into the nature of dark matter.

\end{abstract}

\begingroup
\hypersetup{linkcolor=mahogany} 
\tableofcontents
\endgroup

\section{Introduction}
\label{Intro}

Astrophysical and cosmological observations provide compelling evidence that roughly one quarter of the Universe’s energy budget is non-baryonic dark matter. Flat rotation curves in spiral galaxies first hinted at hidden mass~\cite{RubinFordThonnard}, and the Bullet-Cluster lensing map later separated ordinary matter from the main gravitating centres~\cite{CloweZaritsky}. 
The Planck satellite has since provided a high-precision measurement of the Universe's composition, constraining the cold dark matter density to $\Omega_{\text{DM}}h^2 = 0.1200 \pm 0.0012$ ~\cite{PlanckResults2020}. Yet the Standard Model (SM) contains no suitable particle to fill this role and likewise leaves unresolved the origin of neutrino masses, confirmed by Super-Kamiokande oscillations~\cite{SuperKamiokande1998}; the cosmic baryon asymmetry, whose dynamical generation was framed by Sakharov’s three conditions~\cite{Sakharov}; and the strong-CP problem, elegantly addressed by the Peccei–Quinn mechanism~\cite{PecceiQuinn}. Moreover, next-to-next-to-leading-order analyses place the measured Higgs and top-quark masses close to the boundary where the electroweak vacuum becomes metastable~\cite{Espinosa,ButtazzoStrumia}. Together, these shortcomings make physics beyond the Standard Model unavoidable.

Many comprehensive theories, such as Supersymmetry (SUSY)~\cite{CastanoPiardRamond,Ellwanger} or models with extended gauge groups like 3-3-1~\cite{PisanoPleitez,SanchezVegaGambini,DorschLouziSanchezVegaViglioni,DiasPleitez,MonteroCastellanosSanchezVega,CastellanosSanchezVega,DiasKannikeSanchezVega}, address some of these fundamental issues while naturally providing DM candidates. In these frameworks, however, the DM particle is often just one component of a much larger, more complex structure designed to solve other problems, like the hierarchy problem. While powerful, this approach introduces significant new particle content and dynamics that may not be directly related to the nature of dark matter itself. An alternative, complementary strategy is to pursue minimality: to isolate the DM problem and study it within the simplest possible extensions to the SM. This bottom-up approach allows for a focused investigation of the core interactions that govern DM phenomenology.

The most economical of these minimal extensions is the scalar singlet Higgs-portal model, where a new scalar field, stabilized by a $\mathbb{Z}_2$ symmetry, constitutes the DM~\cite{Silveira:1985rk,Burgess:2000yq,BiswasMajumdar}. While minimal and predictive, this simple scenario is now severely constrained by experimental searches. The viable parameter space has been pushed into two narrow regions: a low-mass window around the Higgs resonance ($M_S\!\simeq\!M_h/2$), which requires very small couplings, and a heavy-mass regime ($M_S\!\gtrsim\!1\;\text{TeV}$)~\cite{ClineScottWeniger,AthronWhite,AthronCornellWild}. These tight constraints naturally motivate the exploration of the next logical step in complexity: multi-component dark matter, where multiple stable particles collectively account for the cosmic density~\cite{BoehmFayetSilk,BargerLangacker,Zurek,ProfumoSigurdsonUbaldi,LiuWuZhou,IvanovKeus,EschKlasenYaguna,Pandey_2018,YagunaZapata2024,DengBotoIvanovSilva}.

A well-motivated and elegant framework for multi-component DM is provided by models with a single, larger discrete symmetry, such as $\mathbb{Z}_N$, which can stabilize multiple particles simultaneously~\cite{Batell,BelangerKannikeRaidal2012,Yaguna_2020}. Such models enrich the phenomenology, allowing for new processes like semi-annihilation and conversions between DM species, which can significantly alter the relic density calculation and observational signatures. This work focuses on a specific class of two-component scalar DM models stabilized by $\mathbb{Z}_{2n}$ symmetries, namely the $\mathbb{Z}_4$ and two $\mathbb{Z}_6$ scenarios originally proposed in Ref.~\cite{YagunaZapata2021}.

However, the viability of these models faces a two-front challenge that demands a new, comprehensive analysis. On the experimental front, the landscape has been transformed by the latest results from the LUX-ZEPLIN (LZ) experiment, which have improved the sensitivity to spin-independent WIMP-nucleon cross-sections by up to a factor of four compared to previous limits~\cite{LZ2024}. This provides the first opportunity to subject these multi-component models to truly stringent direct detection constraints. On the theoretical front, introducing multiple scalar fields makes the model's internal consistency a critical issue. A complete assessment requires a high-scale ``stress test'' to ensure that the scalar potential remains bounded from below (vacuum stability) and that scattering amplitudes respect probability conservation (perturbative unitarity). This approach is standard practice, most notably applied to the SM itself to determine the (meta)stability of the electroweak vacuum~\cite{Espinosa,ButtazzoStrumia} and its lifetime~\cite{AndreassenFrostSchwartz2018}, and is routinely used in other BSM studies~\cite{Sher,GonderingerLimRamseyMusolf,BelangerPukhovYagunaZapata}. We perform this analysis up to the GUT and Planck scales, as they represent the natural energy frontiers where one expects the model to be subsumed by a more fundamental theory, such as a Grand Unified Theory or a theory of quantum gravity.

This paper performs the first combined analysis of these $\mathbb{Z}_{2n}$ models, confronting them simultaneously with the latest, most powerful experimental limits from LZ and a rigorous theoretical evaluation of their high-scale consistency using one-loop Renormalization Group Equations (RGEs). By merging these two fronts, we can delineate a robustly constrained parameter space, offering a more refined and physically consistent picture of these compelling DM scenarios.

To structure our analysis, we begin in Section~\ref{Z2nModels} with a general review of the $\mathbb{Z}_{2n}$ models, highlighting their symmetric properties and scalar sectors. Section~\ref{VacStab} discusses the classical stability conditions of the scalar potential and extends the analysis to the quantum level using a renormalized effective potential to account for radiative corrections. In Section~\ref{PertUnit}, we examine unitarity constraints and their impact on model viability. Section~\ref{DarkMat} explores dark matter phenomenology, focusing on relic abundance constraints and direct detection prospects. In Section~\ref{Results}, we present a comprehensive discussion of our findings and their implications for future experimental and theoretical studies. Finally, Section~\ref{Conc} summarizes our conclusions. Technical details are relegated to the appendices: Appendix~\ref{RGEs} lists the Renormalization Group Equations necessary for the high-scale analysis, while Appendix~\ref{reldens} provides the full coupled Boltzmann equations for the relic density calculation.

\section{The Generalities of the \texorpdfstring{$\mathbb{Z}_{2n}$}{Z2n} Models}
\label{Z2nModels}

The models under consideration extend the SM by incorporating a discrete $\mathbb{Z}_{2n}$ symmetry. In this framework, the scalar sector consists of a complex scalar field, $S_A$, and a real scalar field, $S_B$. The real scalar transforms under $\mathbb{Z}_{2n}$ symmetry as $S_B \to -S_B$, ensuring stability.

The most general Lagrangian for the $\mathbb{Z}_{2n}$ models can be expressed as:
\begin{equation}
    \mathcal{L} = \mathcal{L}_{\text{SM}} + \mathcal{L}_{\text{DM}},
    \label{L2.1}
\end{equation}
where $\mathcal{L}_{\text{SM}}$ represents the SM sector, including $ V_{\text{SM}} = -\mu_H^2 |H|^2 + \lambda_H |H|^4$. The second term, $\mathcal{L}_{\text{DM}}$, represents the dark sector contribution and is given by:
\begin{equation}
    \mathcal{L}_{\text{DM}} = (\partial_\mu S_A^*)(\partial^\mu S_A) + \frac{1}{2} (\partial_\mu S_B)(\partial^\mu S_B) - V(H, S_A, S_B),
\end{equation}
where the scalar potential associated with the extended sector is $V(H, S_A, S_B)=V_{\text{DM}} + V_{\text{int}}$, with each term expressed as:
\begin{align}
    V_{\text{DM}} &= -\mu_A^2 |S_A|^2 + \lambda_A |S_A|^4 - \frac{1}{2} \mu_B^2 S_B^2 + \lambda_B S_B^4, \\
    V_{\text{int}} &= \lambda_{HA} |H|^2 |S_A|^2 + \frac{1}{2} \lambda_{HB} |H|^2 S_B^2 + \lambda_{AB} |S_A|^2 S_B^2.
\end{align}

After electroweak symmetry breaking (EWSB), the Higgs acquires a vacuum expectation value, $\langle H \rangle = \frac{1}{\sqrt{2}}(0,\, v)^\mathsf{T}$, leading to mass terms for the scalars:
\begin{equation}
    M_h^2 = 2\lambda_H v^2, \quad
    M_{S_A}^2 = \frac{1}{2} \lambda_{HA} v^2 - \mu_A^2, \quad
    M_{S_B}^2 = \frac{1}{2} \lambda_{HB} v^2 - \mu_B^2.
\end{equation}

Beyond the general structure described above, the $\mathbb{Z}_4$ and $\mathbb{Z}_6$ models include additional terms, as summarized in Table~\ref{tab:Z4_Z6_terms}. In the $\mathbb{Z}_6$ model, two distinct charge assignments are possible, giving rise to two scenarios, denoted as $\mathbb{Z}_6(23)$ and $\mathbb{Z}_6(13)$. The terms in Table~\ref{tab:Z4_Z6_terms} introduce crucial phenomenological differences between these models. The $\mathbb{Z}_4$ model allows for trilinear and quartic interactions that influence both dark matter phenomenology and the vacuum structure. In contrast, the $\mathbb{Z}_6(23)$ model features a cubic interaction that affects dark matter annihilation channels, while the $\mathbb{Z}_6(13)$ model contains only quartic interactions.

\begin{table}[htbp] 
    \centering
    \begin{tabular}{l l l} 
        \toprule
        \textbf{Model} & \textbf{Model-Specific Interactions} & \textbf{Charge Assignments} \\
        \midrule
        $\mathbb{Z}_4$ & $\dfrac{1}{2} \left(\mu_{S1} S_A^2 S_B + \lambda_{S4} S_A^4\right) + \text{h.c.}$ & $S_A \to \omega_4 S_A, \quad S_B \to \omega_4^2 S_B$ \\ \addlinespace
        $\mathbb{Z}_6(23)$ & $\dfrac{1}{3} \mu_{S2} S_A^3 + \text{h.c.}$ & $S_A \to \omega_6^2 S_A, \quad S_B \to \omega_6^3 S_B$ \\ \addlinespace
        $\mathbb{Z}_6(13)$ & $\dfrac{1}{3} \lambda'_{AB} S_A^3 S_B + \text{h.c.}$ & $S_A \to \omega_6 S_A, \quad S_B \to \omega_6^3 S_B$ \\
        \bottomrule
    \end{tabular}
    \caption{Model-specific interaction terms and charge assignments for the $\mathbb{Z}_{2n}$ models. These terms are in addition to the common potential defined in the main text. Here, $\omega_n = e^{i 2\pi / n}$ represents the $n$-th root of unity. Note that $\omega_4^2 = \omega_6^3 = -1$, which implements the $S_B \to -S_B$ transformation.}
    \label{tab:Z4_Z6_terms}
\end{table}

\section{Classical and Quantum Vacuum Stability}
\label{VacStab}

For a scalar potential to be physically acceptable, it must be bounded from below (BFB); otherwise, the theory becomes unstable due to the absence of a lowest energy state. At tree level, the potential is a polynomial in the scalar fields, and for large field values, its behavior is dominated by the quartic terms, $V_4$. The condition for strong stability is that $V_4$ must remain positive in the large-field limit along any direction in field space.

Our $\mathbb{Z}_{2n}$ models present two distinct structures for $V_4$:
\begin{enumerate}
    \item \textbf{Biquadratic Potentials:} In the $\mathbb{Z}_4$ and $\mathbb{Z}_6(23)$ models, the symmetries restrict the quartic terms to a biquadratic form, which depends only on the squared norms of the fields ($|H|^2, |S_A|^2, S_B^2$). For such potentials, the BFB condition can be elegantly mapped onto the mathematical condition of copositivity for the matrix of quartic couplings.
    \item \textbf{General Potential:} In the $\mathbb{Z}_6(13)$ model, the presence of a term of the form $S_A^3 S_B$ breaks the biquadratic structure. Consequently, the copositivity analysis is not directly applicable, requiring the use of more general stability criteria, such as those derived in \cite{Kannike2016}.
\end{enumerate}

In this section, we first derive the classical stability conditions for each model using the appropriate methodology. We then extend the analysis to the quantum level at one-loop order, adopting the method proposed in \cite{Chataignier}. This approach improves the effective potential by introducing a single, field-dependent renormalization scale, enabling the use of the same stability conditions with renormalization-group-improved quartic couplings.

\subsection{Classical Vacuum Stability}

\subsubsection*{Models with Biquadratic Potentials: The Copositivity Approach ($\mathbb{Z}_4$ and $\mathbb{Z}_6(23)$)}

When the quartic potential, $V_4$, can be written in the form $V_4 = \mathbf{h}^\mathsf{T} \mathbf{\Lambda} \mathbf{h}$, where $\mathbf{h}$ is a vector with non-negative components (in this case, $\mathbf{h}^\mathsf{T} = (|H|^2, |S_A|^2, S_B^2)$) and $\mathbf{\Lambda}$ is a symmetric matrix, the condition $V_4 > 0$ is satisfied if and only if $\mathbf{\Lambda}$ is strictly copositive \cite{Kannike2012, CottleHabetlerLemke}. The presence of cubic terms in the full potential of our models necessitates this strong stability condition. For a $3 \times 3$ symmetric matrix $\mathbf{\Lambda}$, the strict copositivity conditions are given by \cite{Kannike2012,Kaplan}:
\begin{gather}
    \lambda_{ii} > 0, \quad \text{for } i=1,2,3 \nonumber \\
    \bar{\lambda}_{ij} \equiv \lambda_{ij} + \sqrt{\lambda_{ii}\lambda_{jj}} > 0, \quad \text{for } i \neq j \label{eq:copositivity_conditions} \\
    \lambda_{12}\sqrt{\lambda_{33}} + \lambda_{13}\sqrt{\lambda_{22}} + \lambda_{23}\sqrt{\lambda_{11}} + \sqrt{\lambda_{11}\lambda_{22}\lambda_{33}} + \sqrt{2\bar{\lambda}_{12}\bar{\lambda}_{13}\bar{\lambda}_{23}} > 0. \nonumber
\end{gather}
where $\lambda_{ij}$ are the generic elements of the matrix $\mathbf{\Lambda}$.

\medskip
\noindent\textbf{The $\mathbb{Z}_4$ model.} The quartic part of the $\mathbb{Z}_4$ model scalar potential is
\begin{equation}
V_{4}=\lambda_{H}|H|^{4}+\lambda_{A}|S_{A}|^{4}+\lambda_{B}S_{B}^{4}+\lambda_{AB}|S_{A}|^{2}S_{B}^{2}+ \lambda_{HA}|H|^{2}|S_{A}|^{2}+\frac{1}{2}\lambda_{HB}|H|^{2}S_{B}^{2}+\frac{1}{2}\lambda_{S4}\theta|S_{A}|^{4},
\end{equation}
where the orbit parameter $\theta = \hat{S}_{A}^{4}+(\hat{S}_{A}^{*})^{4}$ (with $\hat{S}_{A}\equiv S_{A}/|S_{A}|$) encapsulates the phase dependence of the field $S_A$. As this parameter varies in the interval $\theta \in [-2, 2]$ and the potential depends on it monotonically, the condition for the potential to be bounded from below is most stringently tested at one of the interval's boundaries. This corresponds to the case that minimizes the combination $\lambda_A + \frac{1}{2}\lambda_{S4}\theta$, allowing our analysis to proceed by applying the stability conditions to an effective potential where $\lambda_A$ is replaced by an effective coupling, $\lambda_A - |\lambda_{S4}|$.

The potential can be written as $V_4 = \mathbf{h}^\mathsf{T} \mathbf{\Lambda} \mathbf{h}$, with the symmetric matrix $\mathbf{\Lambda}$ being 
\begin{equation}
    \mathbf{\Lambda}=\begin{pmatrix}
    \lambda_{H} & \frac{\lambda_{HA}}{2} & \frac{\lambda_{HB}}{4} \\
    * & \lambda_{A} - |\lambda_{S4}| & \frac{\lambda_{AB}}{2} \\
    * & * & \lambda_{B}
    \end{pmatrix}.
    \label{eq:Z4_matrix}
\end{equation}
Applying the strict copositivity conditions from Eq.~\eqref{eq:copositivity_conditions} to this matrix yields the full set of tree-level stability constraints for the $\mathbb{Z}_4$ model. The conditions on the diagonal couplings are straightforward:
\begin{equation}
\label{eq:estabilidade_diagonal}
\lambda_H > 0, \quad \lambda_A - |\lambda_{S4}| > 0, \quad \lambda_B > 0.
\end{equation}
The conditions involving the portal couplings, which constrain the off-diagonal elements, include:
\begin{align}
    \overline{\lambda}_{AB} & \equiv \lambda_{AB}+2\sqrt{(\lambda_{A}-|\lambda_{S4}|)\lambda_{B}} > 0, \label{eq:condAB} \\
    \overline{\lambda}_{HA} & \equiv \lambda_{HA}+2\sqrt{\lambda_{H}(\lambda_{A}-|\lambda_{S4}|)} > 0, \label{eq:condHA} \\
    \overline{\lambda}_{HB} & \equiv \lambda_{HB}+2\sqrt{\lambda_{H}\lambda_{B}} > 0. \label{eq:condHB}
\end{align}
Finally, the couplings must also satisfy the third and most complex condition from Eq.~\eqref{eq:copositivity_conditions}, which for the specific couplings of the $\mathbb{Z}_4$ model takes the form:

\begin{equation}
2\lambda_{AB}\sqrt{\lambda_{H}}+2\lambda_{HA}\sqrt{\lambda_{B}}+\lambda_{HB}\sqrt{\lambda_{A}-|\lambda_{S4}|}
+4\sqrt{\lambda_{H}(\lambda_{A}-|\lambda_{S4}|)\lambda_{B}}+\sqrt{2\overline{\lambda}_{HA}\overline{\lambda}_{HB}\overline{\lambda}_{AB}}>0.
\end{equation}

\medskip
\noindent\textbf{The $\mathbb{Z}_6(23)$ model.} The quartic potential for the $\mathbb{Z}_6(23)$ model is free of angular-dependent terms. Its copositivity matrix $\mathbf{\Lambda}$ is therefore identical to that of the $\mathbb{Z}_4$ model (Eq.~\eqref{eq:Z4_matrix}), but with $\lambda_{S4}=0$. As a result, the stability conditions for this model are exactly those of the $\mathbb{Z}_4$ model in the limit $\lambda_{S4} \to 0$.

\subsubsection*{The Model with Non-Biquadratic Structure: $\mathbb{Z}_6(13)$}

The quartic part of the $\mathbb{Z}_6(13)$ scalar potential is
\begin{equation}
    V_{4}=\lambda_{H}|H|^{4}+\lambda_{A}|S_{A}|^{4}+\lambda_{B}S_{B}^{4}+\lambda_{AB}|S_{A}|^{2}S_{B}^{2}+ \lambda_{HA}|H|^{2}|S_{A}|^{2}+\frac{1}{2}\lambda_{HB}|H|^{2}S_{B}^{2}+\frac{1}{3}\lambda_{AB}^{\prime}\theta^{\prime}|S_{A}|^{3}S_{B}.
    \label{V4Z613}
\end{equation}
The term containing $\lambda_{AB}'$ cannot be expressed as a quadratic function of $|S_A|^2$ and $S_B^2$, thus violating the premise of the copositivity analysis. For this case, we turn to the general stability conditions for a potential with a Higgs doublet and two real scalar fields, as detailed in \cite{Kannike2016}. The orbit parameter $\theta' = \hat{S}_A^3 + (\hat{S}_A^*)^3$ varies in the interval $\theta' \in [-2,2]$. The conditions from \cite{Kannike2016} are expressed as a function\footnote{We do not reproduce the lengthy analytic expressions for the vacuum stability conditions here; instead, we refer the reader to the detailed derivations in Ref.~\cite{Kannike2016}. The numerical implementation of these conditions, along with other results from that study, is available in a supplementary Mathematica file associated with the arXiv preprint \cite{Kannike2016}.}
\begin{equation}
    \mathrm{StabCondV12H}\left[\lambda_{40}, \lambda_{31}, \lambda_{22}, \lambda_{13}, \lambda_{04}, \lambda_H, \lambda_{H20}, \lambda_{H11}, \lambda_{H02}\right].
    \label{StabCondV12H}
\end{equation}
After mapping our couplings from Eq.~\eqref{V4Z613} and evaluating at the extrema of $\theta'$, the final condition for the $\mathbb{Z}_6(13)$ model is that the following relation must be satisfied for both choices of sign:
\begin{equation}
    \mathrm{StabCondV12H}\left[\lambda_A, \pm\frac{2}{3}|\lambda_{AB}'|, \lambda_{AB}, 0, \lambda_{B}, \lambda_H, \lambda_{HA}, 0, \frac{1}{2}\lambda_{HB}\right].
\end{equation}

\subsection{Quantum Vacuum Stability}

At the quantum level, radiative corrections modify the classical stability analysis. These are captured by the one-loop effective potential, given by
\begin{equation}
    V_{\text{Eff}}(\mu,\lambda,\phi)= \overbrace{V^{(0)}(\lambda,\phi)}^{\text{Tree level}}\, +\, \overbrace{\frac{\hbar}{64\pi^2}\sum_a n_a m_a^4(\lambda,\phi) \left[\ln \frac{m_a^2(\lambda,\phi)}{\mu^2}-\chi_a\right]}^{\text{One-loop correction (Coleman-Weinberg)}} +\,  \mathcal{O}\left(\hbar^2\right).
    \label{VEff}
\end{equation}
In the expression above, the sum runs over all particles in the theory that acquire a field-dependent mass $m_a(\lambda,\phi)$, and $\mu$ is the renormalization scale. The term $n_a$ represents the number of degrees of freedom for each particle $a$, including a crucial negative sign for fermions which reflects their statistical contribution to the potential. Specifically, $n_a$ is $1$ for a real scalar, $2$ for a complex scalar, $-4$ for a Dirac fermion, and $3$ for a massive vector boson. The constant $\chi_a$ depends on the chosen renormalization scheme. In the standard Modified Minimal Subtraction ($\overline{\text{MS}}$) scheme, used in this work, its value is $\chi_a = 3/2$ for scalars and fermions, and $\chi_a = 5/6$ for vector bosons.

The validity of the perturbative expansion in Eq.~\eqref{VEff} hinges on keeping couplings small ($\lambda < 4\pi$) and avoiding the large logarithmic corrections that arise when particle mass scales differ significantly. To maintain perturbative control across all energy scales, we adopt the robust method of a field-dependent renormalization scale $\mu^*$ \cite{Chataignier}, where the effective potential is matched to the tree-level potential evaluated with running couplings:
\begin{equation}  
    V(\mu^*,\lambda,\phi) = V^{(0)}(\lambda(\mu^*),\phi(\mu^*)).
\end{equation}  
This Renormalization Group (RG) improvement procedure resums the large logarithms, and its key advantage is the direct connection to the classical analysis: the stability conditions derived previously remain valid, but must now be checked using the running couplings, $\lambda(\mu^*)$.

A parameter space point that satisfies the stability conditions with couplings defined at the electroweak scale may therefore fail to do so at higher energies, as the RG evolution can drive the theory into an unstable regime. We use this principle to perform the high-scale consistency ``stress test'' mentioned in the introduction. Rather than calculating a maximum validity scale for each point, we use the RG evolution as a powerful theoretical filter: we test whether a given point in the parameter space remains stable and perturbative all the way up to the GUT and Planck scales. This allows us to first map out the regions of fundamental theoretical viability, which are then explored to find solutions to the dark matter problem.

To implement this approach, we solve the RGEs numerically. The equations were derived and cross-checked using the automated tools \texttt{SARAH} and \texttt{RGBeta} \cite{Staub2014,Thomsen}. A one-loop calculation of the RGEs, as listed in Appendix \ref{RGEs}, is sufficient to capture the dominant quantum effects and provides a robust assessment of the models' high-energy behavior.

\section{Perturbative Unitarity}
\label{PertUnit}

Perturbative unitarity provides a set of high-energy consistency checks that are independent of, yet complementary to, the vacuum stability conditions. While stability ensures that the scalar potential does not lead to a catastrophic collapse of the vacuum, unitarity protects the predictive power of the theory by ensuring that scattering probabilities remain physical at high energies. By analyzing tree-level $2 \to 2$ scalar scattering amplitudes, we derive upper bounds on the quartic couplings, which are crucial for defining the theoretically consistent parameter space of the models.

We employ the standard high-energy approach, where the partial wave expansion of the S-matrix in terms of the total angular momentum $J$ must remain unitary~\cite{LeeQuiggThackerApr1977,LeeQuiggThackerSep1977,MarcianoValenciaWillenbrock,BetreHedriWalker,GoodsellStaub}. In the high-energy limit ($s \to \infty$), and using the Goldstone boson equivalence theorem, the tree-level amplitudes for $2 \to 2$ scalar scattering are governed by the quartic couplings. Unitarity of the $J = 0$ partial wave imposes bounds on the eigenvalues, $a_0^i$, of the scattering matrix. To ensure the validity of the perturbative expansion, we impose the conventional constraint~\cite{MarcianoValenciaWillenbrock,Logan}:
\begin{equation}
|\operatorname{Re} a_0^i\,| \leq \frac{1}{2}, \quad \forall i.
\label{ChosenUnitCond}
\end{equation}
These eigenvalues are calculated from the fourth derivatives of the scalar potential defined by Eq.~\eqref{L2.1} and Table~\ref{tab:Z4_Z6_terms}.

Applying this formalism yields a set of constraints on the quartic couplings. The most stringent bounds arise from the coupled scattering matrix of the scalar states $\{H, S_A, S_B\}$, leading to constraints that are universal to all three $\mathbb{Z}_{2n}$ models considered. These are given by
\begin{align}
\begin{split}
&|\lambda_H| \leq 4\pi, \quad |\lambda_{HA}| \leq 8\pi, \quad |\lambda_{HB}| \leq 8\pi, \quad |x_{1,2,3}| \leq 16\pi,
\end{split}
\end{align}
where $x_{1,2,3}$ are the roots of the cubic equation
\begin{equation}
    x^3 + A x^2 + B x + C = 0,
\end{equation}
with coefficients
\begin{align}
A &= 12\lambda_H + 8\lambda_A + 24\lambda_B, \\
B &= 96\lambda_H\lambda_A + 288\lambda_H\lambda_B + 192\lambda_A\lambda_B - 8\lambda_{AB}^2 - 8\lambda_{HA}^2 - 4\lambda_{HB}^2, \\
C &= 2304\lambda_H\lambda_A\lambda_B - 96\lambda_H\lambda_{AB}^2 - 32\lambda_A\lambda_{HB}^2 - 192\lambda_B\lambda_{HA}^2 + 32\lambda_{AB}\lambda_{HA}\lambda_{HB}.
\end{align}
In addition, some models feature extra constraints arising from their unique interaction terms.
For the $\mathbb{Z}_4$ model, the $\lambda_{S4}$ coupling leads to the additional constraints:
\begin{equation}
\left|\lambda_A \pm 3\lambda_{S_4}\right| \leq 4\pi, \quad |\lambda_{AB}| \leq 4\pi.
\end{equation}
For the $\mathbb{Z}_6(13)$ model, the $\lambda_{AB}'$ term imposes the extra condition:
\begin{equation}
\left|\lambda_A + \lambda_{AB} \pm \sqrt{\left(\lambda_A - \lambda_{AB}\right)^2 + 2\lambda_{AB}'^2}\right| \leq 8\pi.
\end{equation}
The $\mathbb{Z}_6(23)$ model has no such additional terms, and its bounds are identical to those of the $\mathbb{Z}_4$ model in the limit $\lambda_{S4} \to 0$.

Note that setting all couplings associated with the extended scalar sector to zero recovers the well-known Standard Model perturbative unitarity bound, $M_h\lesssim 1\ \mbox{TeV}$~\cite{LeeQuiggThackerApr1977,LeeQuiggThackerSep1977}. However, as with the vacuum stability analysis, a more robust assessment requires checking these unitarity conditions not just with the low-energy values of the couplings, but across all energy scales. Therefore, we apply the same RG-improvement procedure, replacing the classical couplings in our analysis with their running counterparts obtained by solving the RGEs~\cite{GoodsellStaub}. This ensures that the theory remains unitary and predictive up to the GUT and Planck scales, providing a consistent and powerful theoretical filter on the viable parameter space.

\begin{figure}[htbp]
\centering
\begin{subfigure}[t]{0.4\textwidth}
    \includegraphics[width=\textwidth]{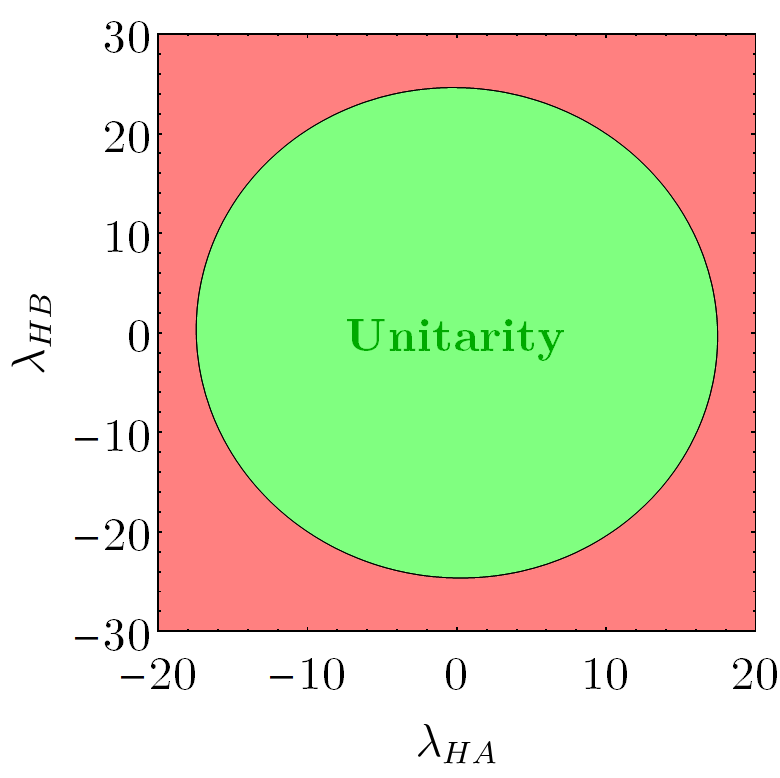}
    \caption{Tree-level (classical) constraints. The allowed parameter space (green) is symmetric under sign changes of the portal couplings ($\lambda_{Hi} \to -\lambda_{Hi}$). While the individual bounds on $\lambda_{HA}$ and $\lambda_{HB}$ are identical, the coupled-channel unitarity analysis treats them asymmetrically, resulting in a non-circular region with a larger allowed range for $\lambda_{HB}$.}
    \label{Z4C1ClasUnit}
\end{subfigure}
\hfill
\begin{subfigure}[t]{0.4\textwidth}
    \includegraphics[width=\textwidth]{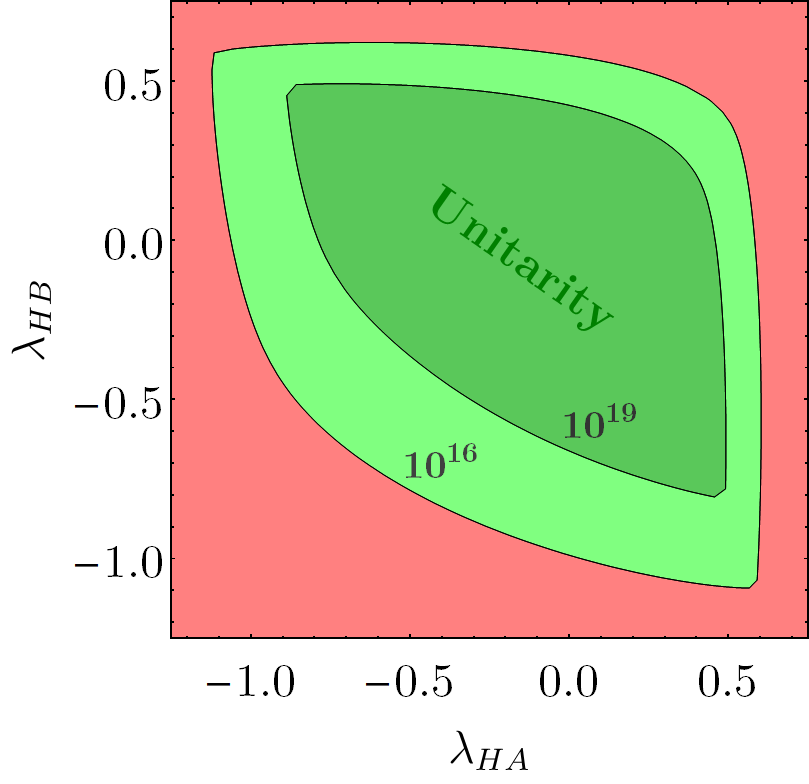}
    \caption{RG-improved constraints. The allowed regions require unitarity to be satisfied up to the GUT scale (light green) and the Planck scale (dark green). The sign-flip symmetry is visibly broken by the one-loop RG evolution.}
    \label{Z4C1QuantUnit}
\end{subfigure}
\caption{Comparison of tree-level and RG-improved perturbative unitarity bounds in the $(\lambda_{HA}, \lambda_{HB})$ plane for the $\mathbb{Z}_4$ model (Scenario 1). The plots highlight the powerful impact of the quantum analysis on the unitarity constraints: while the tree-level calculation allows for couplings in a wide range, roughly $|\lambda_{Hi}| \lesssim 8\pi$, requiring unitarity to hold up to high energies under RG evolution restricts the viable parameter space to the much smaller region where $|\lambda_{Hi}| \lesssim 1$. The initial conditions for the fixed couplings are $\lambda_H \simeq 1/8$, $\lambda_A = 0.03$, $\lambda_B = 0.02$, $\lambda_{AB} = 0.25$, and $\lambda_{S4} = 0.01$. Similar results hold for the other models.}
\label{Z4C1Unit}
\end{figure}

A noteworthy consequence of this RG-improved analysis, readily visible when comparing Fig.~\ref{Z4C1ClasUnit} and Fig.~\ref{Z4C1QuantUnit}, is the breaking of a symmetry present at the classical level. The allowed parameter space at the quantum level is not symmetric under the sign change of the portal couplings $\lambda_{HA}$ and $\lambda_{HB}$. This asymmetry is a direct result of the structure of the one-loop RGEs, detailed in Appendix~\ref{RGEs}, which do not respect this sign-flip symmetry, thus providing a clear example of how quantum corrections can qualitatively alter the landscape of theoretical constraints.

\section{Dark Matter Phenomenology in \texorpdfstring{$\mathbb{Z}_{2n}$}{Z2n} Models}
\label{DarkMat}

Having established the theoretical framework and its high-scale consistency requirements, we now turn to the low-energy phenomenology. The viability of the $\mathbb{Z}_{2n}$ models as a solution to the dark matter problem is tested against two powerful and complementary observational fronts: the cosmological relic abundance, which probes the thermal history of the DM candidates in the early Universe, and direct detection experiments, which search for their interactions with ordinary matter in the present day. This combined analysis, together with the theoretical constraints, forms the basis for our numerical investigation in Section~\ref{Results}.

\subsection{Relic Density}
\label{Relic}

The total DM relic abundance, $\Omega_{\text{DM}}h^2$, is the sum of the contributions from both scalar components, $\Omega_{S_A} h^2$ and $\Omega_{S_B} h^2$. These are determined by numerically solving the coupled Boltzmann equations for the number densities $n_i$:
\begin{equation}
\label{CoupledBoltzmann}
    \frac{dn_{i}}{dt} + 3Hn_{i} = -\sum_{j,k,l} \langle \sigma v \rangle_{ij \to kl} \left( n_i n_j - n_i^{\text{eq}}n_j^{\text{eq}} \frac{n_k n_l}{n_k^{\text{eq}}n_l^{\text{eq}}} \right),
\end{equation}
where $i,j,k,l$ represent either $S_A$, $S_B$, or SM particles. The thermally averaged cross-section, $\langle \sigma v \rangle_{ij \to kl}$, includes all kinematically allowed processes. All three models share a common set of interactions: standard annihilation of each DM component into SM particles (mediated by $\lambda_{HA}$ and $\lambda_{HB}$) and conversion between the two DM species, $S_A S_A^{\dagger} \leftrightarrow S_B S_B$ (mediated by $\lambda_{AB}$).

The distinct phenomenology of each model, however, arises from additional, unique interactions. The $\mathbb{Z}_4$ model features a semi-annihilation channel ($S_A S_A \to S_B h$) mediated by the trilinear coupling $\mu_{S1}$. The $\mathbb{Z}_6(23)$ model also allows for semi-annihilation, but of a different kind ($S_A S_A \to S_A^\dagger h$), driven by the $\mu_{S2}$ term. In contrast, the $\mathbb{Z}_6(13)$ model lacks semi-annihilation but introduces a novel conversion process via the quartic coupling $\lambda'_{AB}$. A comprehensive analysis of these processes is presented in Appendix~\ref{reldens}.

To obtain the relic density, the coupled Boltzmann equations are solved numerically using the public code \texttt{micrOMEGAs}~\cite{micromegas}. The final constraint requires that the total relic abundance matches the value precisely measured by the Planck satellite: $\Omega_{\text{DM}}h^2 = 0.1200 \pm 0.0012$~\cite{PlanckResults2020}. Our analysis considers a model viable if its calculated relic density falls within the $3\sigma$ confidence level of this measurement.

\subsection{Direct Detection}
\label{sec:DirectDetection}

While the relic density probes the thermal history of the dark sector, direct detection experiments offer an orthogonal window, sensitive only to the portal connecting DM to atomic nuclei. These experiments search for rare, low-energy nuclear recoils induced by DM particles scattering elastically off a target material~\cite{Xenon1T, Lux-Zeplin2023, LZ2024, PandaX4T}. In the $\mathbb{Z}_{2n}$ models, the scalar nature of the DM candidates means these interactions are spin-independent (SI) and mediated exclusively by the Higgs portal couplings:
\begin{equation}
\label{eq:HiggPortal}
\mathcal{L}_{\text{int}} \supset -\lambda_{HA} |H|^2 |S_A|^2 - \frac{1}{2}\lambda_{HB} |H|^2 S_B^2.
\end{equation}
The SI WIMP-nucleon cross-section is given by:
\begin{equation}
\label{eq:nucleonXS}
\sigma_{S_i}^{\text{SI, nucleon}} \approx \frac{\lambda_{Hi}^2 f_p^2}{4\pi} \frac{\mu_{ip}^{2} m_{p}^{2}}{M_{h}^{4} M_{S_i}^{2}},
\end{equation}
where $\mu_{ip}$ is the WIMP-proton reduced mass, $m_p$ is the proton mass, and $f_p \approx 0.3$ is the effective WIMP-proton coupling factor. This is the quantity we use to confront our model predictions with experimental bounds.

For a multi-component DM scenario, the experimental limits on $\sigma_{S_i}^{\text{SI, nucleon}}$ are not directly applicable to each particle individually, as the expected event rate scales with each component's fractional contribution to the local DM density, $\xi_{S_i} = \Omega_{S_i}h^2/\Omega_{\text{DM}}h^2$. Following the procedure in~\cite{Betancur}, we therefore impose the combined constraint from direct detection experiments via:
\begin{equation}
\label{eq:res.2}
\frac{\xi_{S_{A}}\sigma_{S_{A}}^{\text{SI,nucleon}}}{\sigma_{\text{limit}}(M_{S_{A}})} + \frac{\xi_{S_{B}}\sigma_{S_{B}}^{\text{SI,nucleon}}}{\sigma_{\text{limit}}(M_{S_{B}})} < 1,
\end{equation}
where $\sigma_{\text{limit}}(M_{S_i})$ represents the most stringent experimental upper limit for a DM particle of mass $M_{S_i}$. For this work, we use the limits from both the initial LZ data release~\cite{Lux-Zeplin2023} and the significantly improved results more recently presented by the collaboration in 2024~\cite{LZ2024}, which allows us to demonstrate the powerful impact of the latest data on the viable parameter space.

\section{Numerical Constraints on \texorpdfstring{$\mathbb{Z}_{2n}$}{Z2n} Dark Matter: Stability, Unitarity, Coupling Perturbativity, and Experiments}
\label{Results}

We now combine the theoretical and phenomenological constraints discussed in the previous sections to explore the viable parameter space of the $\mathbb{Z}_{2n}$ models. For each model, we perform an extensive random scan over the free parameters, with points distributed logarithmically to ensure a thorough exploration of the different energy scales. The most fundamental of these parameters are the DM particle masses, $M_{S_A}$ and $M_{S_B}$, which are universally scanned within the range:
\begin{equation}
    \label{eq:res.3}
    40~\text{GeV} \le M_{S_A}, M_{S_B} \le 2~\text{TeV}.
\end{equation}
This mass range is chosen for two primary reasons. First, it covers the classic WIMP mass scale, which is actively being probed by current and upcoming direct detection experiments and collider searches. Second, it aligns with the mass range studied in the original proposal of these models~\cite{YagunaZapata2021}, allowing for a direct comparison that highlights the impact of our more stringent and comprehensive analysis.

To navigate the large dimensionality of the parameter space, our analysis strategy is guided by the dependencies summarized in Figure~\ref{tabpar}. We first identify theoretically robust regions of the parameter space by performing a broad scan over the quartic couplings and imposing the rigorous criteria of vacuum stability (Section~\ref{VacStab}), perturbative unitarity (Section~\ref{PertUnit}), and coupling perturbativity (Appendix~\ref{RGEs}) up to the GUT and Planck scales.

From within these self-consistent domains, we select two representative benchmark scenarios, detailed in Table~\ref{tab:UnifiedScenariosValues}, to serve as distinct, illustrative starting points for our detailed phenomenological analysis. This approach allows us to focus our scans on the parameters most directly linked to the DM phenomenology, confident that our chosen foundation is theoretically consistent.

\begin{figure}[htbp]
    \centering
    \begin{adjustbox}{max width=\textwidth}
    \begin{tikzpicture}[every node/.style={font=\small}]
        \matrix (m) [
            matrix of nodes,
            nodes={
                draw,
                minimum height=0.8cm,
                anchor=center,
                outer sep=0pt,
            },
            column 1/.style={nodes={text width=2.2cm,align=center}},
            column 2/.style={nodes={fill=blue!10,minimum width=1.5cm}}, 
            column 3/.style={nodes={fill=green!10,minimum width=1.5cm}},
            column 4/.style={nodes={fill=orange!10,minimum width=1.8cm}},
            column 5/.style={nodes={minimum width=1.2cm}},
            column 6/.style={nodes={minimum width=1.4cm}},
            column 7/.style={nodes={minimum width=1.4cm}},
            column sep=0.2cm,
            row sep=0.2cm,
            nodes in empty cells,
            row 1/.style={nodes={font=\bfseries, minimum height=1cm}}
        ]
        {
            Parameter & Relic Density & Direct Detection & Theoretical Constraints & $\mathbb{Z}_4$ & $\mathbb{Z}_6(23)$ & $\mathbb{Z}_6(13)$ \\
            $\lambda_{HA}$ & \textcolor{blue!70!black}{\Large$\bullet$} & \textcolor{green!70!black}{\Large$\bullet$} & \textcolor{orange!70!black}{\Large$\bullet$} & \checkmark & \checkmark & \checkmark \\
            $\lambda_{HB}$ & \textcolor{blue!70!black}{\Large$\bullet$} & \textcolor{green!70!black}{\Large$\bullet$} & \textcolor{orange!70!black}{\Large$\bullet$} & \checkmark & \checkmark & \checkmark \\
            $\lambda_{AB}$ & \textcolor{blue!70!black}{\Large$\bullet$} & {} & \textcolor{orange!70!black}{\Large$\bullet$} & \checkmark & \checkmark & \checkmark \\
            $\mu_{S1}$ & \textcolor{blue!70!black}{\Large$\bullet$} & {} & {} & \checkmark & {} & {} \\
            $\mu_{S2}$ & \textcolor{blue!70!black}{\Large$\bullet$} & {} & {} & {} & \checkmark & {} \\
            $\lambda'_{AB}$ & \textcolor{blue!70!black}{\Large$\bullet$} & {} & \textcolor{orange!70!black}{\Large$\bullet$} & {} & {} & \checkmark \\
            $\lambda_A$ & {} & {} & \textcolor{orange!70!black}{\Large$\bullet$} & \checkmark & \checkmark & \checkmark \\
            $\lambda_B$ & {} & {} & \textcolor{orange!70!black}{\Large$\bullet$} & \checkmark & \checkmark & \checkmark \\
            $\lambda_{S4}$ & {} & {} & \textcolor{orange!70!black}{\Large$\bullet$} & \checkmark & {} & {} \\
        };
    \end{tikzpicture}
    \end{adjustbox}
 \caption{A map of the parameter dependencies for the $\mathbb{Z}_{2n}$ models, which guides our numerical analysis strategy. The colored columns indicate whether a parameter (rows) influences the relic density (blue), direct detection (green), or the theoretical constraints (orange), with a filled circle ($\bullet$) marking a dependency. The final three columns use a checkmark ($\checkmark$) to show which parameters are present in each specific model. Note that some parameters are unique to a single model: $\mu_{S1}$ and $\lambda_{S4}$ ($\mathbb{Z}_4$); $\mu_{S2}$ ($\mathbb{Z}_6(23)$); and $\lambda'_{AB}$ ($\mathbb{Z}_6(13)$).}

    \label{tabpar}
\end{figure}

\begin{table}[htbp]
\centering
\renewcommand{\arraystretch}{1.2} 
\begin{tabularx}{\textwidth}{@{} l c c *{3}{>{\centering\arraybackslash}X} @{}}
\toprule
 &  &  & \multicolumn{3}{c}{Model Applicability} \\
 \cmidrule(lr){4-6}
Coupling & Scenario 1 & Scenario 2 & $\mathbb{Z}_4$ & $\mathbb{Z}_6(23)$ & $\mathbb{Z}_6(13)$   \\
\midrule
$\lambda_{A}$ & 0.03 & 0.07 & \checkmark & \checkmark & \checkmark  \\
$\lambda_{B}$ & 0.02 & 0.035 & \checkmark & \checkmark & \checkmark  \\
$\lambda_{AB}$ & 0.25 & -0.05 & \checkmark & \checkmark & \checkmark  \\
$\lambda_{S4}$ & 0.01 &  0.05  & \checkmark &  —  &  —   \\
$\lambda'_{AB}$ & 0.01 &  0.05  &  —  &  — & \checkmark  \\
\bottomrule
\end{tabularx}
\caption{Benchmark values for the fixed quartic couplings that define the two main scenarios explored in our numerical analysis. These scenarios are constructed to be theoretically robust, satisfying the constraints of vacuum stability, perturbative unitarity, and coupling perturbativity up to the GUT and Planck scales. The checkmarks ($\checkmark$) in the `Model Applicability' columns indicate which parameters are relevant for each of the three $\mathbb{Z}_{2n}$ models.}
\label{tab:UnifiedScenariosValues}
\end{table}

\subsection{\texorpdfstring{$\mathbb{Z}_{4}$}{Z4} Model}
\label{ResultsZ4Model}

\begin{figure}[htbp]
    \centering
    \begin{subfigure}{0.4\textwidth}
        \centering
        \includegraphics[width=\textwidth]{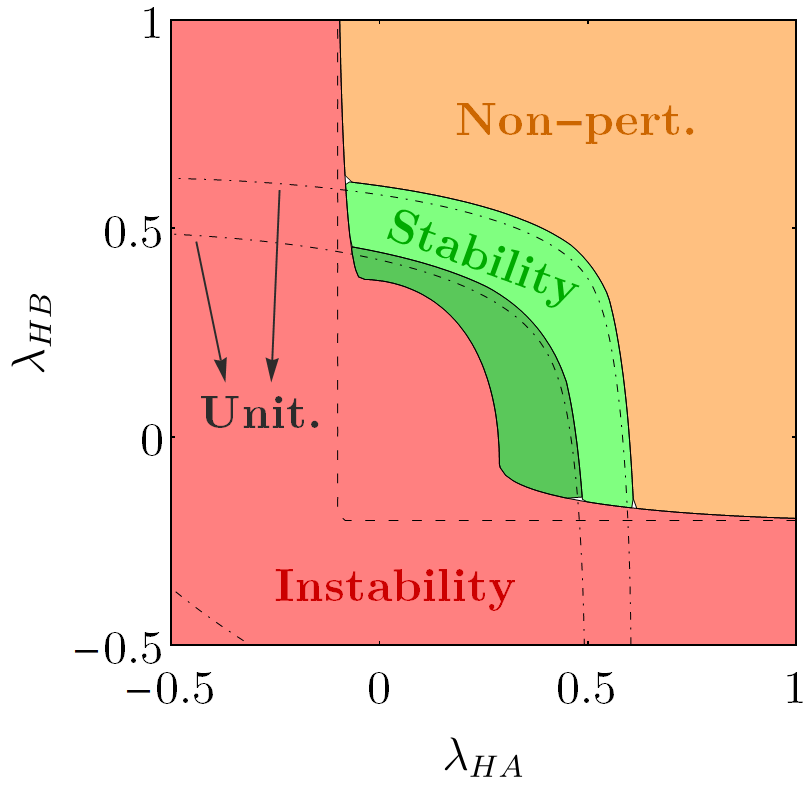}
        \caption{Scenario 1.}
        \label{fig:Z4Cen1}
    \end{subfigure}
    \hfill
    \begin{subfigure}{0.4\textwidth}
        \centering
        \includegraphics[width=\textwidth]{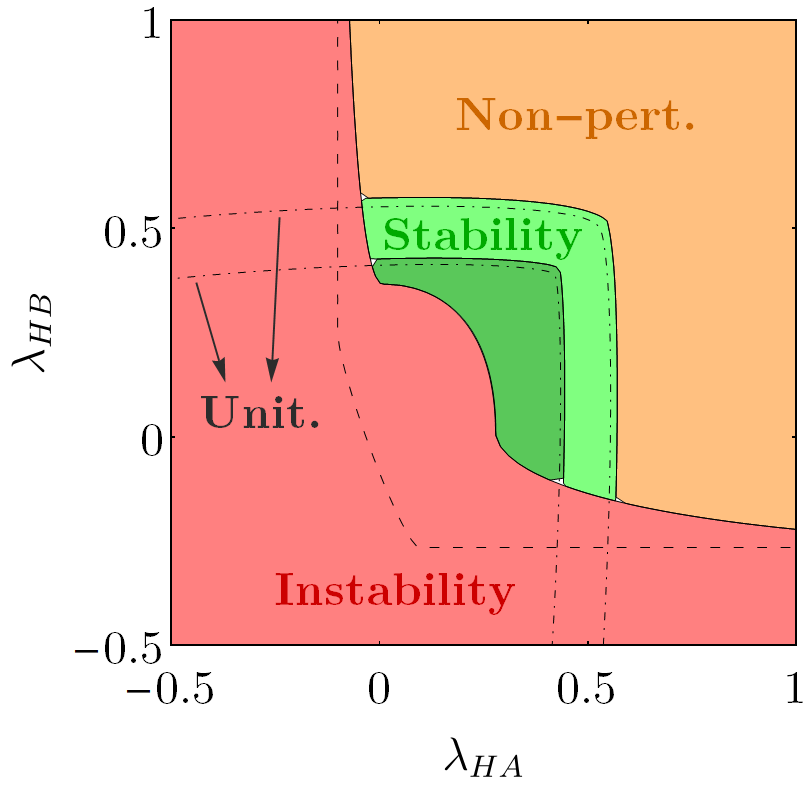}
        \caption{Scenario 2.}
        \label{fig:Z4Cen2}
    \end{subfigure}
   \caption{Theoretically consistent regions in the $(\lambda_{HA}, \lambda_{HB})$ plane for the two benchmark scenarios of the $\mathbb{Z}_4$ model (see Table~\ref{tab:UnifiedScenariosValues}). The color coding denotes regions that are: stable up to the Planck scale (dark green); stable only up to the GUT scale (light green); unstable at a high scale (red); or non-perturbative (orange). The dashed and dot-dashed lines indicate the boundaries of classical stability and perturbative unitarity, respectively.}
\label{fig:Z4cases}
\end{figure}

We begin our detailed numerical analysis with the $\mathbb{Z}_4$ model. The analysis proceeds in two distinct stages. First, we establish the theoretically allowed landscape for the model, independent of any dark matter considerations. For each of the two benchmark scenarios defined in Table~\ref{tab:UnifiedScenariosValues}, we scan the Higgs portal couplings ($\lambda_{HA}, \lambda_{HB}$) and apply the high-scale constraints of vacuum stability, perturbative unitarity, and coupling perturbativity. The results of this purely theoretical analysis are presented in Figure~\ref{fig:Z4cases}.

These plots serve as the foundational maps for our phenomenological study, illustrating the impact of high-scale consistency requirements on the parameter space. The color coding is defined as follows.

The theoretically consistent regions are presented as nested sets. The \textbf{dark green} area identifies the most robust parameter space, where the model remains stable and perturbative up to the Planck scale. The \textbf{light green} area, in turn, represents points that satisfy these criteria up to the GUT scale, but not all the way to the Planck scale.

Conversely, the \textbf{red} region marks parameter points where the scalar potential becomes unbounded from below. This instability may be present already at the classical level or be induced by the RG evolution at a higher energy scale. We emphasize that our analysis tests for absolute stability; points within this red region are not rigorously excluded, as we do not assess the potential for a cosmologically long-lived metastable vacuum. The \textbf{orange} regions are ruled out by loss of perturbativity, where at least one coupling exceeds the non-perturbative limit of $|\lambda_i| \ge 4\pi$.

The \textbf{dashed} line delineates the region of the parameter space where the scalar potential is bounded from below at the tree level, thus ensuring vacuum stability. Additionally, the \textbf{dot-dashed} lines represent the constraints derived from perturbative unitarity up to the Planck/GUT scale. It is notable that these unitarity bounds largely coincide with and reinforce the stability limits, particularly for large field values, highlighting the internal consistency of the model.

Having established these theoretically allowed regions, we now investigate which points can simultaneously satisfy the dark matter constraints. To do this, we perform our extensive random scan over the free parameters governing the DM phenomenology. The Higgs portal couplings are scanned within the range $-1 \le \lambda_{HA}, \lambda_{HB} \le 1$, while the trilinear coupling $\mu_{S1}$ is varied over:
\begin{equation}
    100\ \text{GeV} \leq \mu_{S1} \leq 10\ \text{TeV}. \label{eq:res.4}
\end{equation}

The DM masses, $M_{S_{A}}$ and $M_{S_{B}}$, are scanned within the interval defined in Eq.~\eqref{eq:res.3}. For the mass ordering, we impose the hierarchy $M_{S_{A}} < M_{S_{B}}$. In the opposite mass hierarchy ($M_{S_{B}} < M_{S_{A}}$), the overall depletion is still controlled by the $S_{A}$-mediated semi-annihilation. Flipping the ordering mainly changes which component saturates $\Omega_{\rm DM}$—the lighter $S_{B}$ typically accounts for nearly the entire abundance—without enlarging the allowed $(\lambda_{HA},\lambda_{HB})$ window or qualitatively altering the semi-annihilation and direct-detection patterns discussed above. Crucially, across the phenomenologically relevant sub-TeV parameter space the scenario develops a strong near-degeneracy, $M_{S_{B}} \simeq M_{S_{A}}$, which limits the prospects for distinct experimental signatures between the two dark matter components. For these reasons, we concentrate our discussion on the case $M_{S_{A}} < M_{S_{B}}$. Furthermore, to ensure the stability of both DM components, the mass of the heavier particle must satisfy the kinematic constraint $M_{S_{B}} < 2M_{S_{A}}$, which forbids the decay $S_{B} \rightarrow S_{A} S_{A}$ mediated by the trilinear coupling $\mu_{S1}$.

The resulting points are then analyzed and classified for visualization in the subsequent figures. We first require all points to satisfy the Planck relic density constraint. Then, to illustrate the impact of the remaining bounds, we apply the following coloring scheme. The stringent direct detection limits from LZ 2024 are applied first. Points that violate this constraint are categorized in two ways to provide more detail: configurations where \textit{both} DM candidates would be individually ruled out by the LZ limit are colored black. In contrast, configurations that are also excluded by the combined constraint (Eq.~\ref{eq:res.2}), but where only one of the two components would individually exceed the limit, are colored gray. We refer to this latter case as `partially excluded' for clarity, although both black and gray points are ultimately ruled out phenomenologically. Finally, the points that are \textit{not} excluded by direct detection are colored according to their high-scale theoretical properties. This is determined by the location of their $(\lambda_{HA}, \lambda_{HB})$ values in the foundational maps of Figure~\ref{fig:Z4cases}. Specifically, points whose coordinates fall within either the instability (\textbf{red}) or non-perturbative (\textbf{orange}) domains of the theoretical map are collectively colored \textbf{red} in the subsequent phenomenological plots.

Conversely, points that lie in the consistent regions are colored according to their maximum scale of validity: \textbf{light green} for those stable up to the GUT scale, and \textbf{dark green} for those stable up to the Planck scale. This classification scheme provides a clear visual representation of the interplay between all constraints.

\begin{figure}[htbp]
    \centering

    \begin{subfigure}{1.0\textwidth}
        \centering
        \begin{minipage}{0.45\textwidth}
            \centering
            \includegraphics[width=\textwidth]{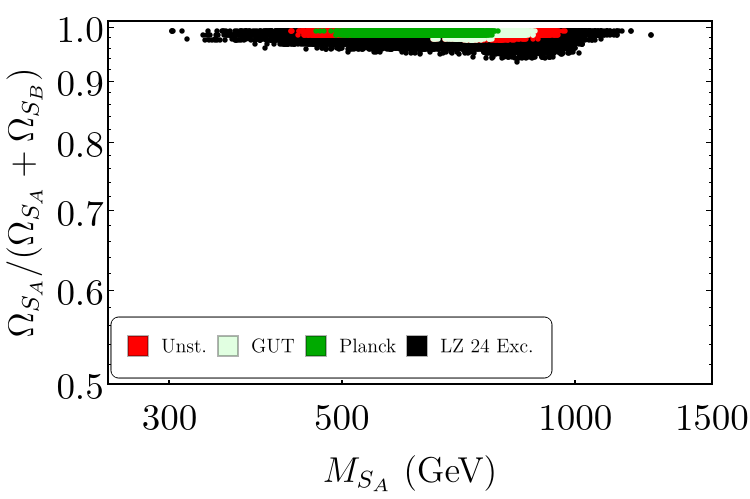}
        \end{minipage}
        \hfill
        \begin{minipage}{0.45\textwidth}
            \centering
            \includegraphics[width=\textwidth]{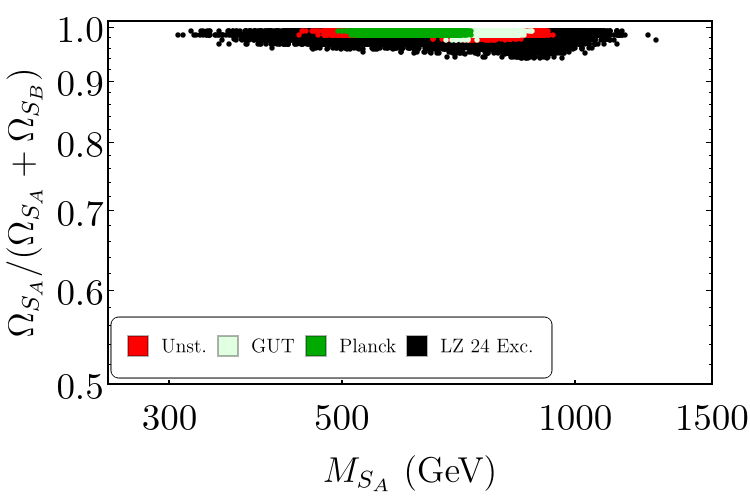}
        \end{minipage}
        \caption{Fractional contribution of the lighter DM component, $S_A$, to the total relic density versus its mass, $M_{S_A}$, for Scenario 1 (left) and Scenario 2 (right). The plots demonstrate that $S_A$ is the dominant DM component across the viable parameter space. The color coding classifies the points according to our filtering scheme: points excluded by LZ (2024) are shown in black. Points that survive this cut are then colored by their high-scale stability: unstable (red), stable up to the GUT scale (light green), and stable up to the Planck scale (dark green).}
        \label{fig:Z4Graph1}
    \end{subfigure}

    \vspace{1em}

    \begin{subfigure}{1.0\textwidth}
        \centering
        \begin{minipage}{0.45\textwidth}
            \centering
            \includegraphics[width=\textwidth]{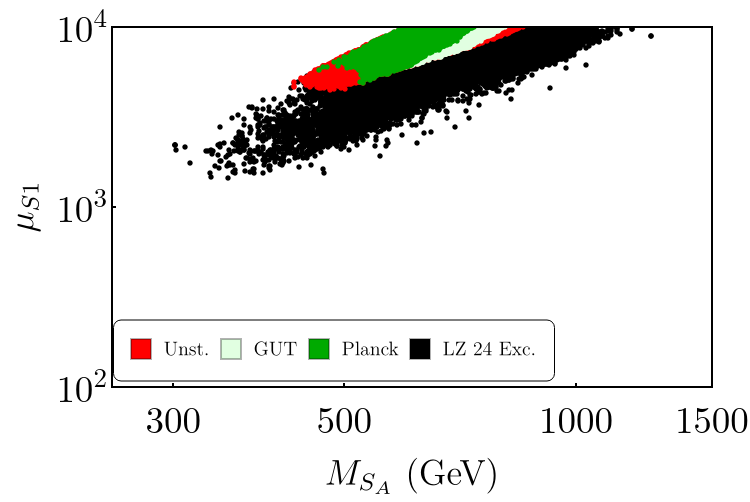}
        \end{minipage}
        \hfill
        \begin{minipage}{0.45\textwidth}
            \centering
            \includegraphics[width=\textwidth]{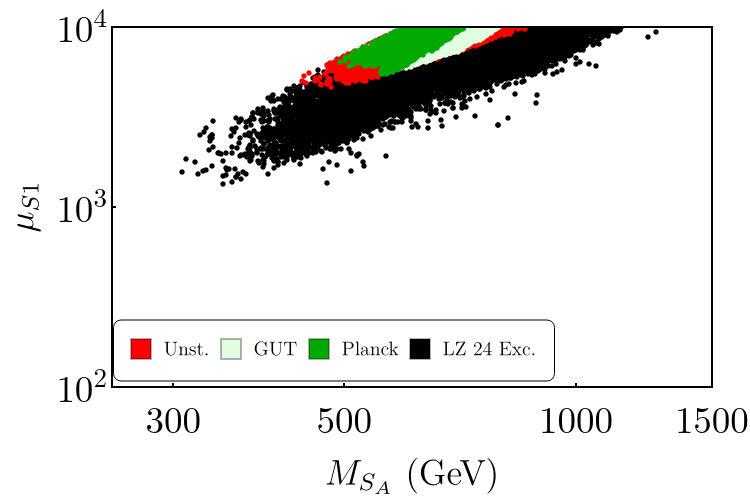}
        \end{minipage}
        \caption{The correlation between the DM mass $M_{S_A}$ and the trilinear coupling $\mu_{S1}$ for Scenario 1 (left) and Scenario 2 (right). This plot illustrates the crucial role of semi-annihilation in setting the relic abundance: larger values of $\mu_{S1}$ are required to efficiently deplete the DM density for heavier candidates, thus satisfying the Planck constraint. The color coding for the points is the same as in Fig.~\ref{fig:Z4Graph1}.}
        \label{fig:Z4Graph2}
    \end{subfigure}

    \vspace{1em}

    \begin{subfigure}{1.0\textwidth}
        \centering
        \begin{minipage}{0.45\textwidth}
            \centering
            \includegraphics[width=\textwidth]{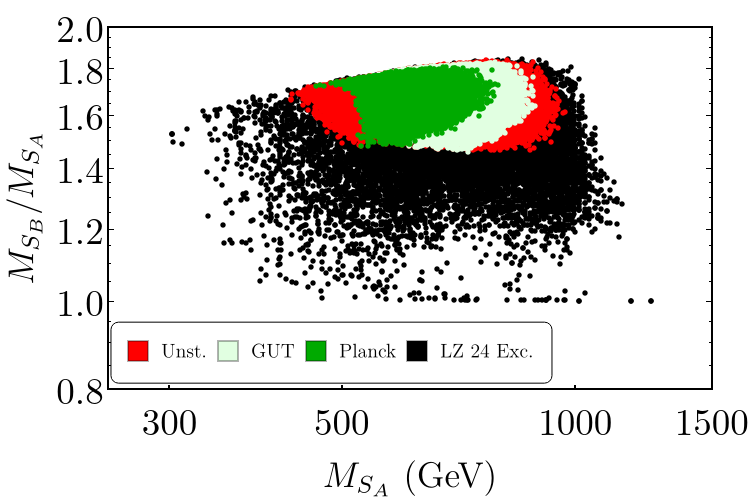}
        \end{minipage}
        \hfill
        \begin{minipage}{0.45\textwidth}
            \centering
            \includegraphics[width=\textwidth]{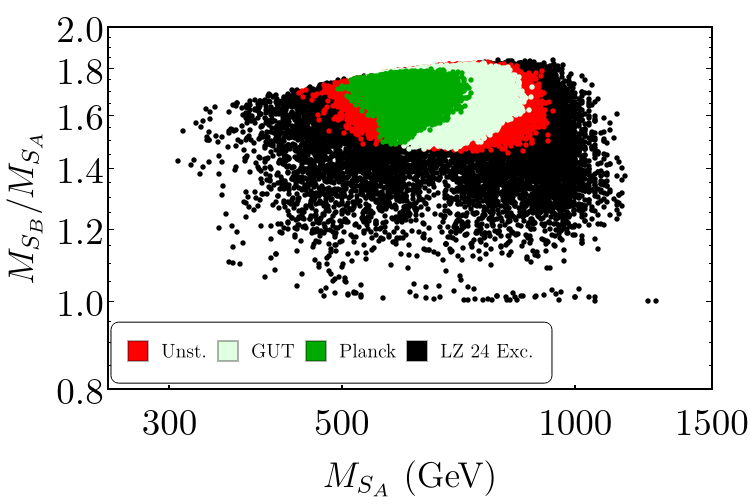}
        \end{minipage}
        \caption{The resulting mass spectrum of the viable DM candidates, showing the mass ratio $M_{S_B}/M_{S_A}$ as a function of $M_{S_A}$ for Scenario 1 (left) and Scenario 2 (right). The plot demonstrates that all points satisfying the full set of constraints exhibit a non-degenerate mass hierarchy ($M_{S_B}/M_{S_A} > 1$). This clear mass splitting is a key phenomenological feature of the model, allowing for distinct experimental signatures for each component. The color coding is the same as in Fig.~\ref{fig:Z4Graph1}.}
        \label{fig:Z4Graph3}
    \end{subfigure}
\caption{Phenomenology of the viable parameter space for the $\mathbb{Z}_4$ model. The three panels collectively illustrate how the semi-annihilation process, governed by the trilinear coupling $\mu_{S1}$, shapes the model's characteristics. This dynamic leads to the lighter component, $S_A$, dominating the relic density and ensures a non-degenerate mass spectrum, as detailed in the subcaptions.}
    \label{fig:diagramZ_allx}
\end{figure}

Applying the classification scheme outlined above, we now examine the phenomenological properties of the points that survive all constraints. Figure~\ref{fig:Z4Graph1} shows the fractional contribution of $S_A$ to the total relic density as a function of its mass. As expected from the imposed mass hierarchy ($M_{S_A} < M_{S_B}$), the lighter component, $S_A$, is the dominant contributor to the relic density across most of the viable parameter space. The efficiency of the annihilation processes that set this abundance is strongly influenced by the trilinear coupling $\mu_{S1}$, whose crucial role is explored in Figure~\ref{fig:Z4Graph2}. We observe that the allowed values for this coupling are concentrated in the upper part of the scanned range, typically between $\mu_{S1} \approx 5$ TeV and $10$ TeV, particularly for masses in the $M_{S_A} \approx 500 - 750$ GeV window. This correlation arises because a stronger semi-annihilation rate is required to efficiently deplete the relic density for heavier DM particles, thereby keeping the total abundance consistent with the Planck observation. Conversely, achieving lower values for $\mu_{S1}$ while satisfying the relic density constraint would necessitate higher values for the Higgs portal couplings ($\lambda_{HA}, \lambda_{HB}$). However, such increased portal couplings would lead to a spin-independent cross-section that violates the stringent direct detection limits and would simultaneously risk driving the theory into the non-perturbative regions at high scales (as shown in Figure~\ref{fig:Z4cases}).

Finally, this interplay between the component masses and the semi-annihilation rate results in the mass spectrum shown in Figure~\ref{fig:Z4Graph3}. The plot of the mass ratio $M_{S_B}/M_{S_A}$ versus $M_{S_A}$ confirms that the allowed points consistently exhibit a non-degenerate mass spectrum ($M_{S_B}/M_{S_A} > 1$). This is a desirable feature, as it allows for a clear distinction between the two DM candidates in potential future signals, for example, in collider searches or indirect detection.

\begin{figure}[htbp]
    \centering
    \begin{minipage}{0.45\textwidth}
        \centering
        \includegraphics[width=\textwidth]{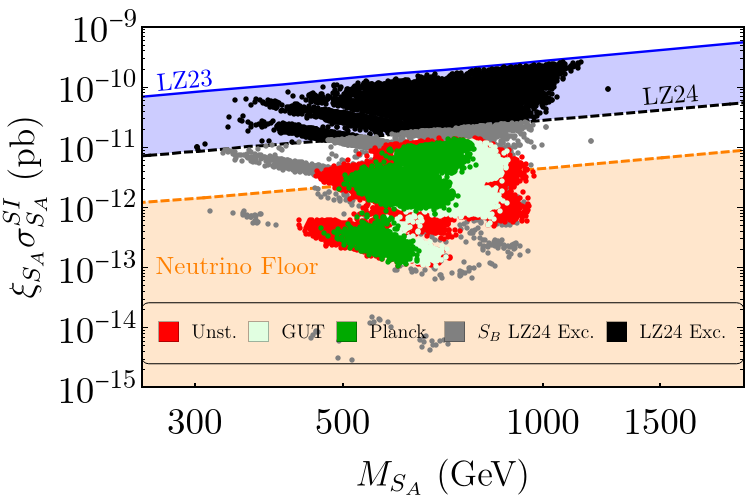}
    \end{minipage}
    \hfill
    \begin{minipage}{0.45\textwidth}
        \centering
        \includegraphics[width=\textwidth]{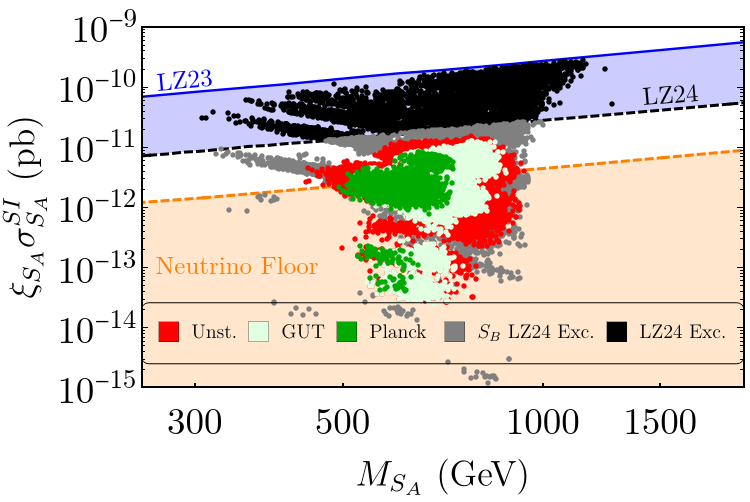}
    \end{minipage}

    \begin{minipage}{0.45\textwidth}
        \centering
        \includegraphics[width=\textwidth]{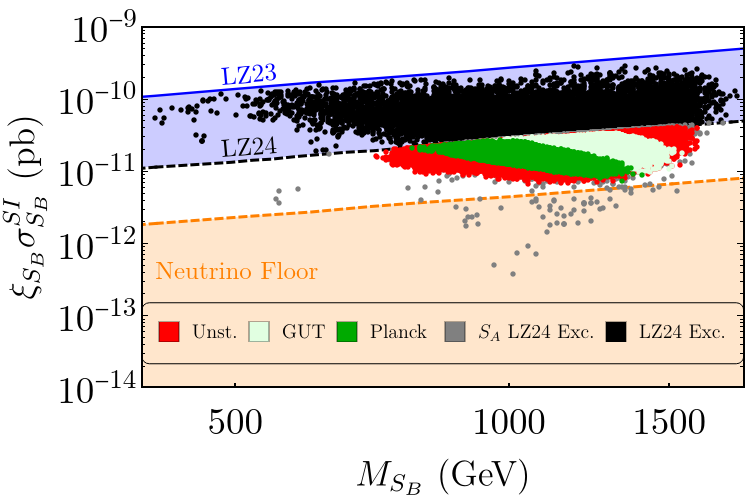}
    \end{minipage}
    \hfill
    \begin{minipage}{0.45\textwidth}
        \centering
        \includegraphics[width=\textwidth]{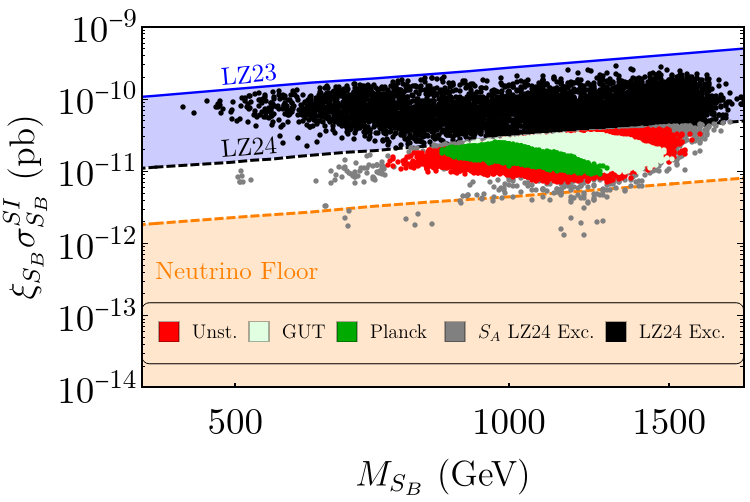}
    \end{minipage}
    \caption{Direct detection prospects for the $\mathbb{Z}_4$ model, showing the rescaled spin-independent cross-section versus mass for each DM component ($S_A$, top; $S_B$, bottom) in Scenario 1 (left) and Scenario 2 (right). The plots highlight the powerful constraining impact of the latest LZ (2024) data (dashed black line) compared to the previous LZ (2023) limit (solid blue line). Points are colored by their status: excluded by LZ (2024) (black and gray), or, if viable, by their high-scale theoretical stability (red, light/dark green). A significant population of viable points remains above the neutrino floor (dashed orange line), indicating strong prospects for future discovery.}
    \label{fig:Z4Graph4A5}
\end{figure}

While the combined direct detection limit was already used as a crucial filter, it is instructive to now analyze the individual contributions of each DM candidate in more detail. Figure~\ref{fig:Z4Graph4A5} provides this detailed view, presenting the rescaled spin-independent WIMP-nucleon scattering cross-section, $\xi_{S_i}\sigma_{S_i}^{\text{SI}}$, as a function of mass for both components. The plot starkly illustrates the impact of the latest LZ (2024) data~\cite{LZ2024}, which carves out a significant portion of the parameter space previously allowed by the LZ (2023) results~\cite{Lux-Zeplin2023}. It is here that the distinction between the excluded points, defined in our classification scheme, becomes visually apparent: black points represent configurations where both DM candidates are individually ruled out, whereas the gray points show the `partially excluded' scenarios where only one component would survive the individual limit, but the pair is nonetheless excluded by the combined constraint of Eq.~\eqref{eq:res.2}. Despite these stringent bounds, a robust population of viable points remains, with a considerable fraction lying above the neutrino floor~\cite{NeutrinoFloor}, indicating strong potential for future discovery.

\begin{figure}[htbp]
    \centering
    \begin{subfigure}{0.45\textwidth}
        \centering
        \includegraphics[width=\textwidth]{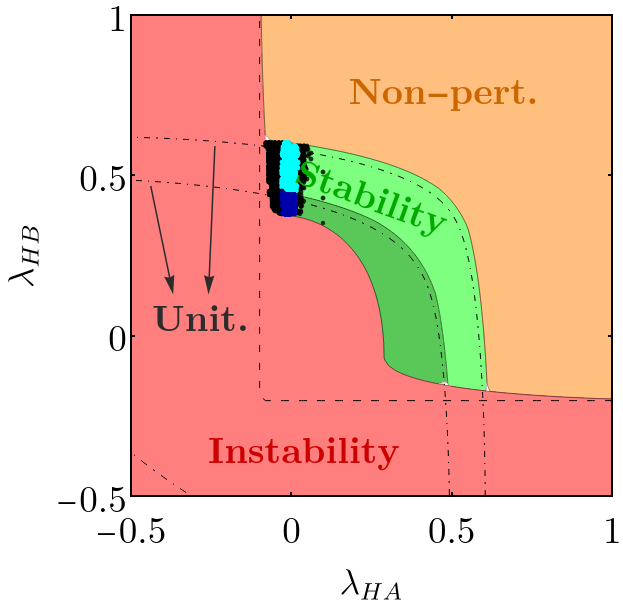}
        \caption{Scenario 1.}
    \end{subfigure}
    \hfill
    \begin{subfigure}{0.45\textwidth}
        \centering
        \includegraphics[width=\textwidth]{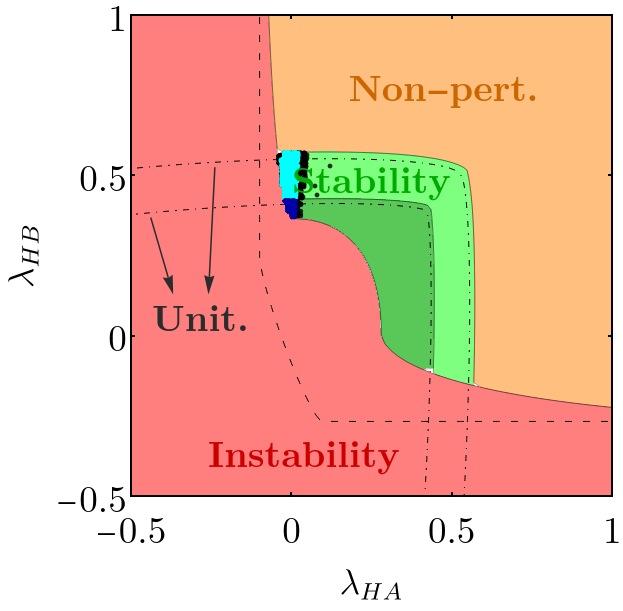}
        \caption{Scenario 2.}
    \end{subfigure}
    \caption{The final, fully constrained parameter space for the $\mathbb{Z}_4$ model in the $(\lambda_{HA}, \lambda_{HB})$ plane for Scenario 1 (left) and Scenario 2 (right). The points from our numerical scan that satisfy both relic density and direct detection constraints are overlaid on the theoretically consistent regions (background colors) from Fig.~\ref{fig:Z4cases}. Viable points are colored dark/light blue, corresponding to their stability up to the Planck/GUT scale, while points excluded by LZ (2024) are shown in black. The result visually demonstrates how only a small fraction of the theoretically viable parameter space survives the stringent phenomenological tests.}
\label{fig:UnifiedVenn}
    \label{fig:Z4Graph6}
\end{figure}

To synthesize all constraints into a single, comprehensive picture, we map these phenomenologically viable points back onto the fundamental theoretical plane of the Higgs portal couplings. The result, shown in Figure~\ref{fig:Z4Graph6}, reveals the final, fully constrained parameter space for the $\mathbb{Z}_4$ model. It is immediately evident that only a small subset of the theoretically consistent regions, originally identified in Figure~\ref{fig:Z4cases}, can simultaneously satisfy the dark matter relic density and direct detection bounds. These surviving points appear as dense clusters, primarily within the regions stable up to the Planck scale (Dark Blue). This visually demonstrates the powerful synergy between theoretical consistency and experimental data in delineating the true parameter space of the model.

A key takeaway from the detailed analysis of the $\mathbb{Z}_4$ model is the qualitative similarity between the results for Scenario 1 and Scenario 2. While the precise boundaries and the total volume of the allowed parameter space are sensitive to the specific values of the fixed quartic couplings, the overall shape of the theoretically consistent regions in the $(\lambda_{HA}, \lambda_{HB})$ plane and the resulting phenomenological trends remain robust. This observation informs our presentation strategy for the subsequent models. To avoid unnecessary repetition, we will often focus the graphical representation of results on a single, representative scenario. It is crucial to note, however, that our full numerical analysis was performed for both scenarios in all models to confirm that this robustness holds and that no new qualitative features emerge from the unique interactions present in the $\mathbb{Z}_6$ models.

\subsection{\texorpdfstring{$\mathbb{Z}_{6}(23)$}{Z6(23)} model}
\label{ResultsZ623Model}

We now apply our analytical framework to the $\mathbb{Z}_6(23)$ model. As detailed in Figure~\ref{tabpar}, its parameter dependencies are similar to the $\mathbb{Z}_4$ case, but with one crucial difference in the interaction driving the dark matter phenomenology. The trilinear coupling $\mu_{S1}$, which couples $S_A$ and $S_B$, is replaced by a cubic self-interaction term for the complex scalar, $\mu_{S2} S_A^3$ (see Section~\ref{Z2nModels} for details). This substitution also simplifies the theoretical constraints, as the model does not contain the $\lambda_{S4}$ coupling.

This new trilinear term, $\mu_{S2}$, still primarily affects the relic density through a semi-annihilation channel, leaving it unconstrained by direct detection or the high-scale theoretical tests. We therefore follow the same strategy as before, treating it as a free parameter in our numerical scan and varying it within the range:
\begin{equation}
\label{eq:res.6}
100 \ \text{GeV} \leq \mu_{S2} \leq 10 \ \text{TeV}.
\end{equation}

\begin{figure}[htbp]
    \centering
    \includegraphics[width=0.45\textwidth]{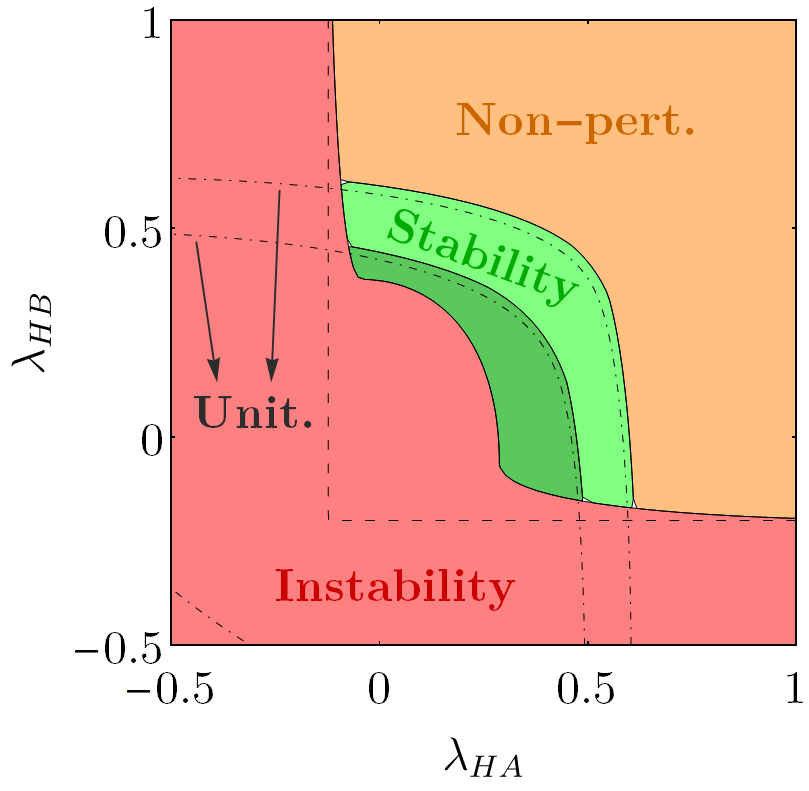}
    \caption{Theoretically consistent regions in the $(\lambda_{HA}, \lambda_{HB})$ plane for a representative benchmark (Scenario 1) of the $\mathbb{Z}_6(23)$ model. As discussed in the text, the results for Scenario 2 are qualitatively similar. The color coding and line style definitions are the same as those used for the $\mathbb{Z}_4$ model in Fig.~\ref{fig:Z4cases}.}
    \label{fig:Z623cases}
\end{figure}

To ensure theoretical consistency, we follow the same strategy as in the $\mathbb{Z}_4$ analysis, fixing the quartic couplings $\lambda_{A}$, $\lambda_{B}$, and $\lambda_{AB}$ to the benchmark values presented in Table~\ref{tab:UnifiedScenariosValues} (with $\lambda_{S4} = 0$). As established in the previous section, the qualitative features of the theoretically allowed regions are robust for both scenarios. Therefore, to avoid repetition, we focus our graphical representation on Scenario 1, noting that our full analysis confirmed this similarity. The resulting stability regions in the $(\lambda_{HA}, \lambda_{HB})$ plane for this representative scenario are shown in Fig.~\ref{fig:Z623cases} and are, as expected, qualitatively similar to those of the $\mathbb{Z}_4$ model (compare with Fig.~\ref{fig:Z4cases}).

We now proceed with the phenomenological scan for our representative benchmark, Scenario 1. The free parameters in our scan are the DM particle masses, $M_{S_A}$ and $M_{S_B}$, the trilinear coupling $\mu_{S2}$, and the Higgs portal couplings, $\lambda_{HA}$ and $\lambda_{HB}$. While the $\mathbb{Z}_6(23)$ model itself does not impose a strict mass hierarchy, for this analysis we maintain the condition $M_{S_A} < M_{S_B} < 2M_{S_A}$, analogous to the $\mathbb{Z}_4$ case, to facilitate comparison. The Higgs portal couplings are varied within the range: $-1 \leq \lambda_{HA}, \lambda_{HB} \leq 1$.

The DM masses and the parameter $\mu_{S2}$ are scanned within the intervals previously defined in Eq.~\eqref{eq:res.3} and Eq.~\eqref{eq:res.6}, respectively.

The results of our scan for the $\mathbb{Z}_6(23)$ model highlight a similar interplay between theoretical and experimental constraints as seen in the $\mathbb{Z}_4$ case. A key phenomenological feature we investigate is whether this model's multi-component nature and semi-annihilation channel can likewise open up a viable parameter space for dark matter candidates with masses below the multi-TeV scale, a region typically excluded for simpler real or complex singlet scalar models~\cite{Silveira:1985rk,McDonald:1993ex,Burgess:2000yq}. The following analysis will show that while such a low-mass region does exist, it is heavily constrained by the combination of the latest direct detection data and our high-scale theoretical tests, leading to a distinct viable parameter space.

\begin{figure}[htbp]
    \centering

    \begin{subfigure}{1.0\textwidth}
        \centering
        \begin{minipage}{0.45\textwidth}
            \centering
            \includegraphics[width=\textwidth]{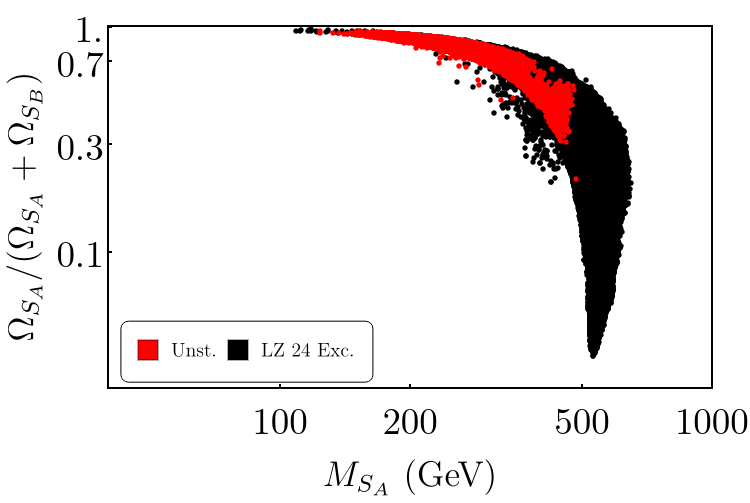}
        \end{minipage}
        \caption{Fractional $S_A$ contribution to the total relic density for the $\mathbb{Z}_6(23)$ model. While $S_A$ remains the dominant component for the points not excluded by direct detection (typically accounting for over 70\% of the total density), this model allows for a more significant sub-dominant contribution from $S_B$ compared to the near-total dominance of $S_A$ in the $\mathbb{Z}_4$ scenario. The color coding indicates points excluded by LZ (2024) (black) and those that are phenomenologically viable but theoretically unstable (red).}
        \label{fig:Z623Graph1}
    \end{subfigure}

    \vspace{1em}

    \begin{subfigure}{1.0\textwidth}
        \centering
        \begin{minipage}{0.45\textwidth}
            \centering
            \includegraphics[width=\textwidth]{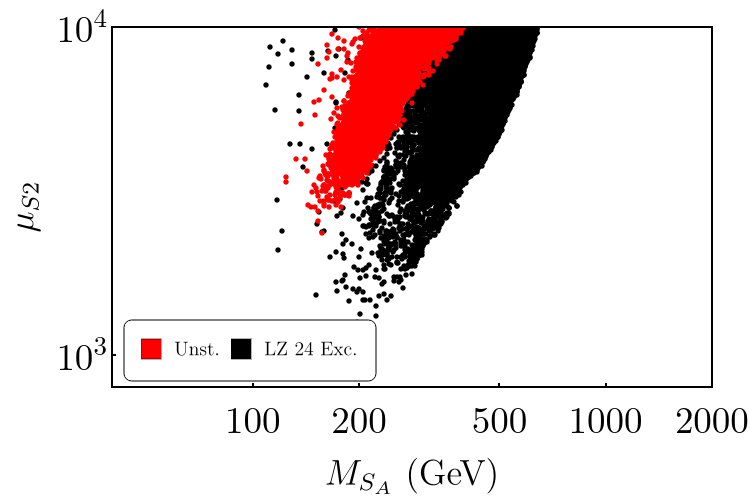}
        \end{minipage}
            \caption{The correlation between the DM mass $M_{S_A}$ and the trilinear coupling $\mu_{S2}$. A larger coupling is required to achieve the correct relic density for heavier DM candidates via semi-annihilation. The color coding is the same as in Fig.~\ref{fig:Z623Graph1}.} 
        \label{fig:Z623Graph3}
    \end{subfigure}

    \vspace{1em}

 \begin{subfigure}{1.0\textwidth}
        \centering
        \begin{minipage}{0.45\textwidth}
            \centering
            \includegraphics[width=\textwidth]{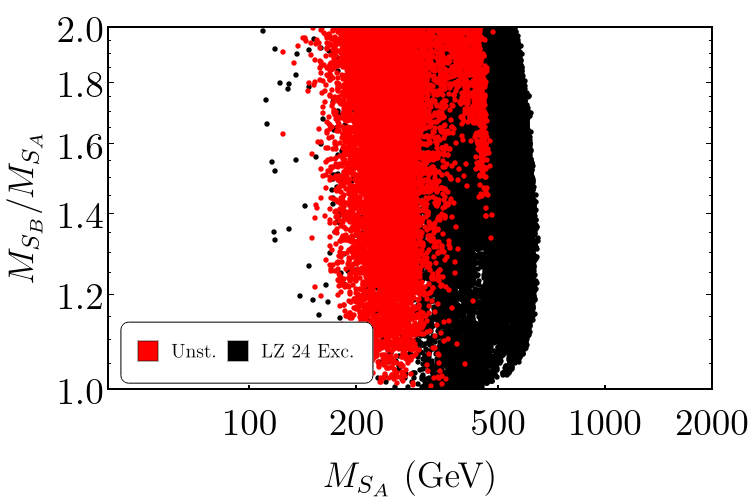}
        \end{minipage}
 \caption{The mass spectrum for the $\mathbb{Z}_6(23)$ model, showing the mass ratio $M_{S_B}/M_{S_A}$ as a function of $M_{S_A}$. Although the mass hierarchy ($M_{S_B} > M_{S_A}$) was imposed in our scan, the plot confirms that the model is viable across a wide range of mass splittings, not just in a degenerate limit. This is a phenomenologically desirable feature, allowing for distinct experimental signatures for each component. The color coding is the same as in Fig.~\ref{fig:Z623Graph1}.}
        \label{fig:Z623Graph2}
    \end{subfigure}
\caption{Phenomenology of the $\mathbb{Z}_6(23)$ model, illustrating a key finding of our analysis: a strong tension between low-energy constraints and high-scale theory. The panels show the characteristics of the parameter space satisfying relic density and direct detection, a region that we find becomes entirely unstable under RG evolution. The relic density dynamics are governed by the semi-annihilation coupling $\mu_{S2}$.}
    \label{fig:diagramZ623_allx}
    
\end{figure}

Our analysis for the $\mathbb{Z}_6(23)$ model reveals a crucial insight: a strong tension exists between its low-energy phenomenology and its high-scale theoretical consistency. Unlike the $\mathbb{Z}_4$ scenario, we find that the entire parameter space region satisfying the relic density and direct detection bounds becomes theoretically unstable under RG evolution. While this finding points to the need for a UV completion to render the model fully consistent, it is instructive to analyze the characteristics of this phenomenologically-driven region, which are detailed in Figures~\ref{fig:Z623Graph1} through~\ref{fig:Z623Graph2}. Figure~\ref{fig:Z623Graph1} shows that while the lighter component, $S_A$, still constitutes the bulk of the relic abundance (typically over 70\%), the model allows for a more significant sub-dominant contribution from $S_B$ compared to the $\mathbb{Z}_4$ case. This dynamic is still governed by the semi-annihilation process, driven by the trilinear coupling $\mu_{S2}$. As shown in Figure~\ref{fig:Z623Graph3}, a strong correlation between $M_{S_A}$ and $\mu_{S2}$ persists. Finally, Figure~\ref{fig:Z623Graph2} shows the resulting mass spectrum. It is important to note that while the mass hierarchy ($M_{S_B} > M_{S_A}$) was imposed in our scan, the model proves to be phenomenologically viable across a significant range of mass splittings, not just in a degenerate limit.

\begin{figure}[htbp]
    \centering
    \begin{subfigure}{0.45\textwidth}
        \centering
        \includegraphics[width=\textwidth]{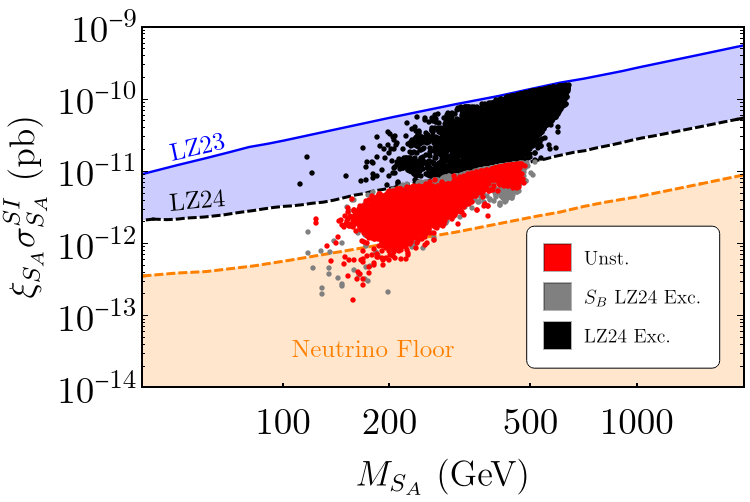} 
        \caption{$\xi_{S_A}\sigma_{S_A}^{\text{SI}}$ vs. $M_{S_A}$ for the $\mathbb{Z}_6(23)$ model (Scenario 1).}
        \label{fig:Z623Graph4}
    \end{subfigure}
    \hfill
    \begin{subfigure}{0.45\textwidth}
        \centering
        \includegraphics[width=\textwidth]{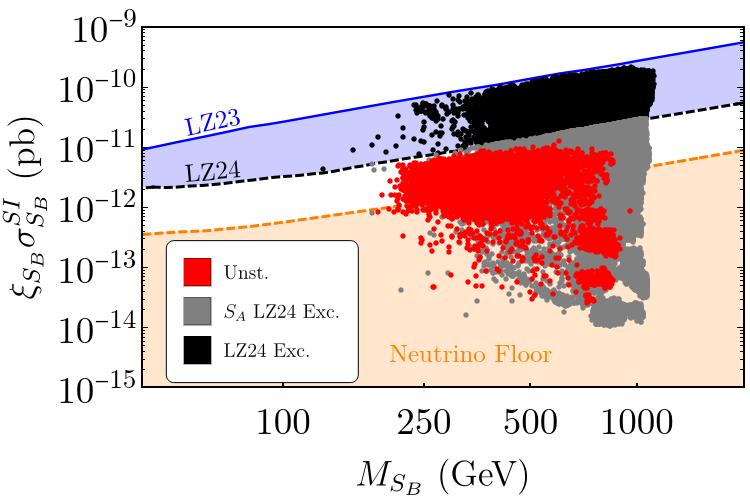} 
        \caption{$\xi_{S_B}\sigma_{S_B}^{\text{SI}}$ vs. $M_{S_B}$ for the $\mathbb{Z}_6(23)$ model (Scenario 1).}
        \label{fig:Z623Graph5}
    \end{subfigure}
\caption{Direct detection prospects for the $\mathbb{Z}_6(23)$ model, showing the rescaled cross-section for $S_A$ (a) and $S_B$ (b). This plot provides a direct visualization of the model's central tension: all points that satisfy the stringent LZ (2024) limit (dashed black line) are theoretically unstable (red). Points that fall within the region constrained by LZ (2024) are shown in black and gray. For context, the previous LZ (2023) limit (solid blue) and the neutrino floor (dashed orange) are also displayed.}
    \label{fig:Z623Graph4A5}
\end{figure}

The direct detection prospects, detailed in Figure~\ref{fig:Z623Graph4A5}, provide the final and most direct visualization of the central tension within the $\mathbb{Z}_6(23)$ model. The plots starkly confirm our central finding: the only points that survive the stringent LZ (2024) direct detection limits~\cite{LZ2024} are those colored red, signifying that the entire phenomenologically viable region is theoretically unstable. The distinction between the excluded points is also clear, with fully excluded (black) and `partially excluded' (gray) configurations populating the parameter space. Notably, a significant fraction of this phenomenologically-driven, albeit theoretically unstable, region lies above the neutrino floor~\cite{NeutrinoFloor}, suggesting that future experiments could continue to probe this interesting theoretical landscape.

\begin{figure}[htbp]
    \centering
\includegraphics[width=0.45\textwidth]{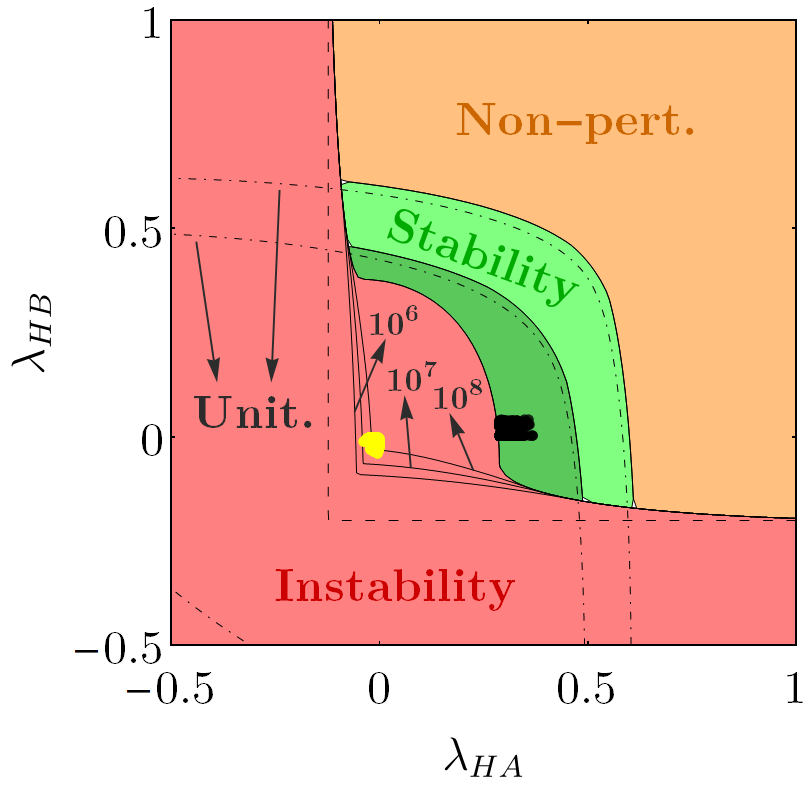}
\caption{A synthesis of all constraints for the $\mathbb{Z}_6(23)$ model, mapping the phenomenologically-driven results onto the fundamental theoretical landscape of the $(\lambda_{HA}, \lambda_{HB})$ plane. The background colors show the theoretical consistency domains when the model is assumed valid up to the GUT and Planck scales (green regions), while the points represent the parameter space favored by dark matter phenomenology (yellow) and the portion of the stable space already excluded by LZ data (black). The plot illustrates the model's tension at high energies, as the yellow points lie outside the GUT-scale stability region. However, the solid contour lines reveal a crucial insight by showing the boundaries of absolute stability for lower maximum energy scales: $10^8$, $10^7$, and $10^6$ GeV. The key result is immediately apparent: the entire phenomenologically viable region becomes theoretically consistent provided the model is interpreted as an effective theory with a cutoff scale of $\Lambda_{\text{max}} \lesssim 10^6$ GeV. This reframes the model as a viable scenario that points towards the existence of new physics at or below the PeV scale.}
\label{fig:Z623Graph6}
\end{figure}

Finally, we synthesize all constraints in the fundamental plane of the Higgs portal couplings in Figure~\ref{fig:Z623Graph6}. This plot vividly illustrates the tension between theory and phenomenology for the $\mathbb{Z}_6(23)$ model. The points satisfying the DM constraints are overlaid on the theoretical landscape, revealing a clear separation: the black points are theoretically viable (lying within the stable green regions) but are excluded by LZ direct detection limits. Conversely, the yellow points are phenomenologically viable but fall within the theoretically unstable (red) regions. The two clusters arise solely from the interplay between relic-density and direct-detection constraints, which select disjoint regions in the $(\lambda_{HA},\lambda_{HB})$ plane consistent with the required semi-annihilation efficiency and the upper bound on the spin-independent cross section. It is particularly noteworthy that these phenomenologically-driven points also cluster remarkably close to the classical stability boundary, indicating that the model is pushed to the very edge of theoretical consistency to satisfy the DM constraints. This interplay demonstrates that, within our framework, no parameter space for the $\mathbb{Z}_6(23)$ model simultaneously satisfies all theoretical and experimental requirements.

The apparent disconnect between the phenomenologically favored region and high-scale theoretical consistency in the $\mathbb{Z}_6(23)$ model motivates a more nuanced question: up to what energy scale can this model be considered a valid, self-consistent theory? To investigate the threshold at which this tension emerges, we have re-evaluated the stability constraints not at the distant GUT and Planck scales, but at various intermediate energy scales, as illustrated in Figure~\ref{fig:Z623Graph6}.

Our analysis reveals a crucial finding. The entire parameter space that satisfies the dark matter phenomenological requirements (the yellow points in the figure) remains comfortably within the domain of absolute stability, provided the model is considered valid up to a maximum energy scale of $\Lambda_{\text{max}} \lesssim 10^6$ GeV. As this cutoff is raised, for instance to $10^8$ GeV, the RGE running drives the couplings into the unstable region, re-establishing the tension. This result offers a compelling reinterpretation: rather than being fundamentally flawed, the minimal $\mathbb{Z}_6(23)$ model should be viewed as a consistent low-energy effective field theory. This implies that for this model to be a viable explanation for dark matter, it requires a UV completion with new physics appearing at or below the PeV scale ($\sim 10^6$ GeV) to ensure the ultimate stability of the theory. The phenomenologically viable region (yellow points) becomes unstable under the RGE flow. The portal couplings, limited by relic-density requirements and LZ bounds, are too weak to compensate the SM-driven negative running of $\lambda_H$ at high scales.

\subsection{\texorpdfstring{$\mathbb{Z}_6(13)$}{Z6(13)} model}
\label{ResultsZ613Model}

Our analysis now turns to the $\mathbb{Z}_6(13)$ model, a framework that, while theoretically interesting, is known to face significant phenomenological challenges. As first noted in Ref.~\cite{YagunaZapata2021}, the model struggles to simultaneously satisfy relic density and direct detection constraints. The root of this difficulty lies in its unique interaction structure. Unlike the previous two models, it lacks a trilinear coupling for efficient semi-annihilation. Instead, its dark sector dynamics are governed by the quartic interaction $\lambda'_{AB} S_A^3 S_B$, as shown in Figure~\ref{tabpar}.

This structural difference is crucial: the same coupling that controls number density conversion also impacts the high-scale theoretical constraints of vacuum stability and perturbative unitarity. Given this pivotal dual role, our analysis strategy for this model must differ from the previous cases. To fully explore the impact of this key parameter, we now treat $\lambda'_{AB}$ as a free variable in our scan, varying it within the range $-1 \le \lambda'_{AB} \le 1$. Meanwhile, the other dark sector quartic couplings ($\lambda_A$, $\lambda_B$, and $\lambda_{AB}$) are fixed to the benchmark values of our representative Scenario 1 (see Table~\ref{tab:UnifiedScenariosValues}). The remaining free parameters in our scan are the DM masses and the Higgs portal couplings, for which we follow the same methodology as the previous models.

\begin{figure}[htbp]
    \centering
    \includegraphics[width=0.4\textwidth]{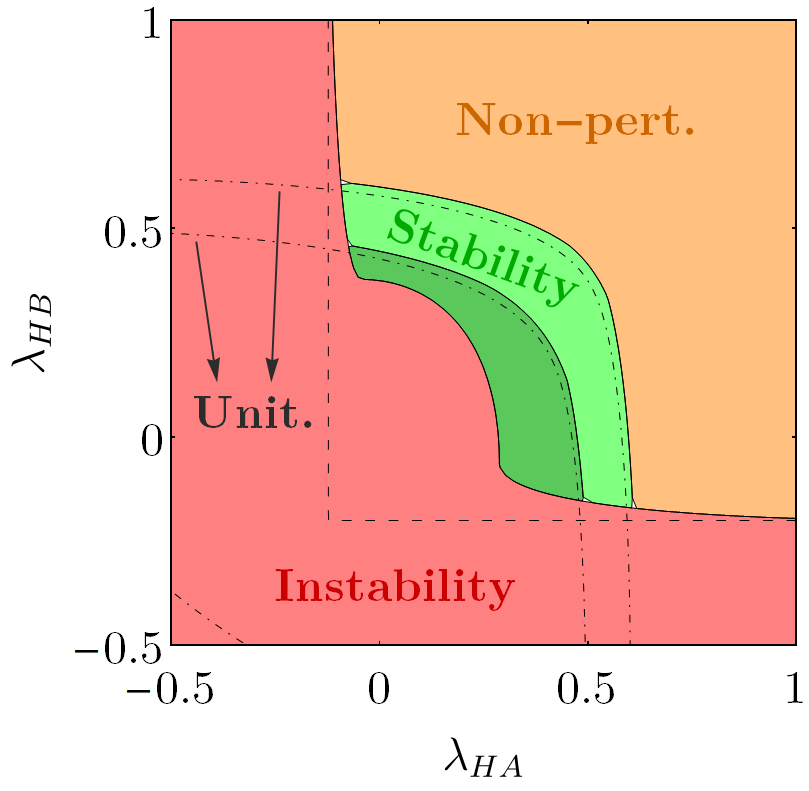}
    \caption{Theoretically consistent regions in the $(\lambda_{HA}, \lambda_{HB})$ plane for a representative benchmark (Scenario 1) of the $\mathbb{Z}_6(13)$ model. This map is generated for a fixed initial value of the key coupling, $\lambda'_{AB}(M_t)=0.01$. As discussed in the text, the results for Scenario 2 are qualitatively similar. The color coding and line style definitions are the same as those used in Fig.~\ref{fig:Z4cases}.}
    \label{fig:Z613cases}
\end{figure}

As in the previous sections, we begin our analysis by establishing the theoretically allowed landscape for the $\mathbb{Z}_6(13)$ model. A key difference, however, is that the high-scale constraints in this model now also depend on the value of the new quartic coupling, $\lambda'_{AB}$. To construct a representative map of the theoretically consistent regions, we fix this coupling to its benchmark value for Scenario 1, $\lambda'_{AB}(M_t) = 0.01$. It is crucial to note that this value is not held constant throughout the analysis; it evolves with energy according to its RGE (see Appendix~\ref{RGEs}), thus dynamically influencing the stability and unitarity boundaries at high scales. The resulting analysis for the $(\lambda_{HA}, \lambda_{HB})$ plane is presented in Figure~\ref{fig:Z613cases}, which, as anticipated, shows regions that are qualitatively similar to those of the other models.

\begin{figure}[htbp]
    \centering

    \begin{subfigure}{1.0\textwidth}
        \centering
        \begin{minipage}{0.45\textwidth}
            \centering
            \includegraphics[width=\textwidth]{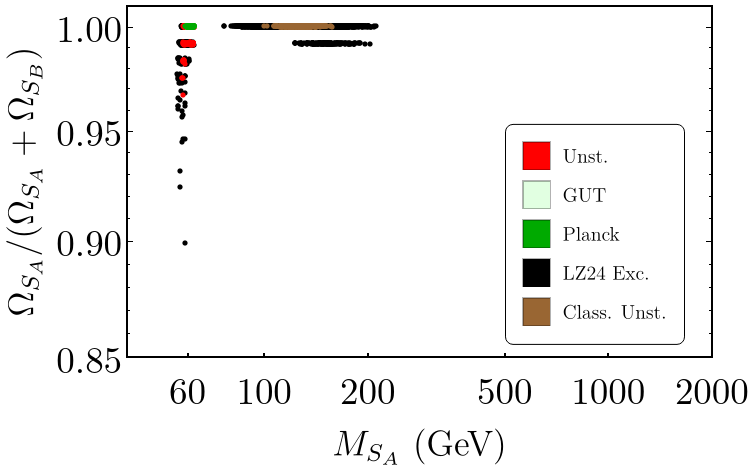}
        \end{minipage}
          \caption{Fractional contribution of $S_A$ to the total relic density, confirming its dominance across the entire viable parameter space. The color coding distinguishes points by their experimental status and theoretical stability (classical, GUT, or Planck scale).} 
        \label{fig:Z613Graph1}
    \end{subfigure}

    \vspace{1em}

    \begin{subfigure}{1.0\textwidth}
        \centering
        \begin{minipage}{0.45\textwidth}
            \centering
            \includegraphics[width=\textwidth]{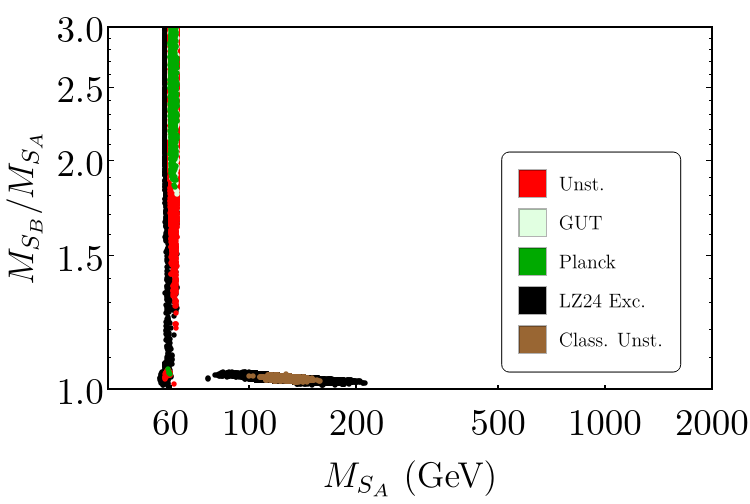}
        \end{minipage}
      \caption{The resulting mass spectrum. A clear mass splitting is observed in the viable Higgs resonance region, while a strong degeneracy ($M_{S_B} \approx M_{S_A}$) appears outside of it, limiting the potential for distinct experimental signatures in that regime. The color coding is the same as in Fig.~\ref{fig:Z613Graph1}.}
        \label{fig:Z613Graph2}
    \end{subfigure}
     \caption{Phenomenological properties of the highly restricted parameter space for the $\mathbb{Z}_6(13)$ model. The panels illustrate that the model is viable almost exclusively in the Higgs resonance region, where the lighter component, $S_A$, dominates the relic density and a non-degenerate mass spectrum is achieved.}
    \label{fig:diagramZ613_allx}
\end{figure}

\begin{figure}[htbp]
    \centering
    \begin{subfigure}{0.45\textwidth}
        \centering
        \includegraphics[width=\textwidth]{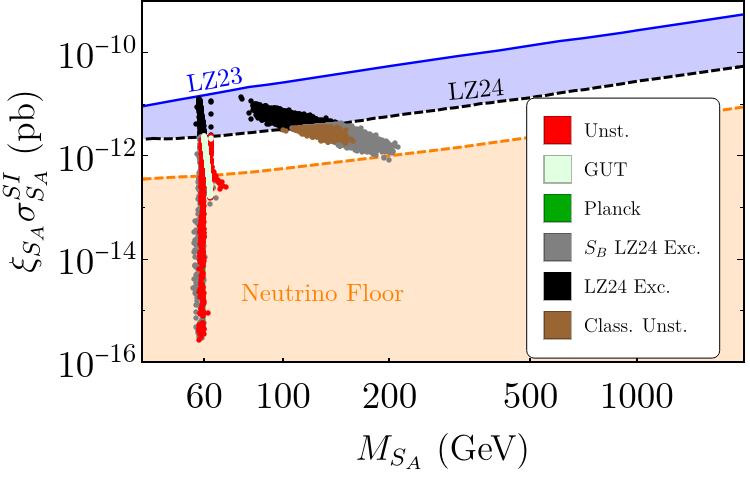} 
        \caption{}
        \label{fig:Z613Graph3}
    \end{subfigure}
    \hfill
    \begin{subfigure}{0.45\textwidth}
        \centering
        \includegraphics[width=\textwidth]{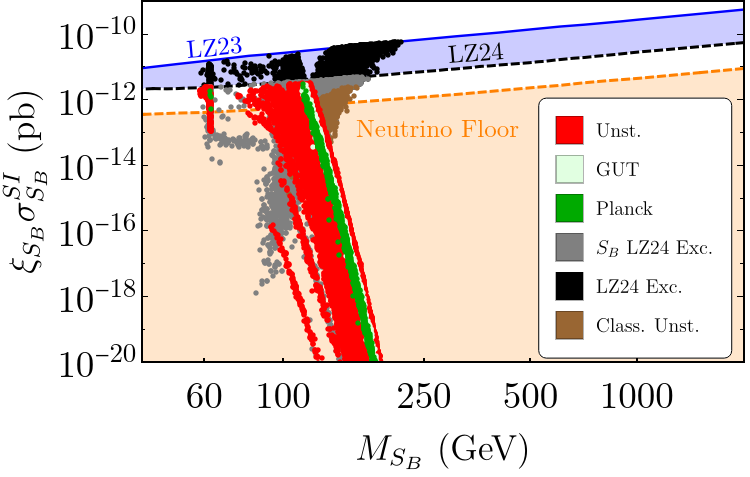} 
          \caption{}
        \label{fig:Z613Graph4}
    \end{subfigure}
\caption{The direct detection signatures for the $\mathbb{Z}_6(13)$ model, showing the rescaled cross-section for $S_A$ (left) and $S_B$ (right). The plot reinforces the model's highly constrained nature, with the few viable points (green) clustering almost exclusively in the Higgs resonance region. The vast majority of the parameter space is challenged by theoretical instabilities (brown, red) or falls within the region constrained by the stringent LZ (2024) limit (black, gray). For context, the previous LZ (2023) limit~\cite{Lux-Zeplin2023} and the neutrino floor~\cite{NeutrinoFloor} are also displayed.}
    \label{fig:Z613Graph3A4}
\end{figure}

Our comprehensive scan of the $\mathbb{Z}_6(13)$ model confirms the significant phenomenological challenges previously identified in the literature. Even with the crucial coupling $\lambda'_{AB}$ treated as a free parameter, our analysis reveals that the viable parameter space is highly restricted. As shown in Figures~\ref{fig:Z613Graph1} through \ref{fig:Z613Graph4}, the points that satisfy the full set of theoretical and experimental constraints cluster almost exclusively in the well-known Higgs resonance region ($M_{S_A} \approx M_h/2$). This result underscores the difficulty of achieving the correct relic density in a model that lacks an efficient semi-annihilation channel. Furthermore, we find that outside this narrow resonance window, virtually all parameter points tend to be classically unstable. To visualize this complex interplay of constraints in the subsequent figures, we adopt the following classification scheme: priority is given to direct detection, with points excluded by the LZ (2024) experiment colored black. Points not excluded by direct detection are then colored according to their theoretical stability: brown for classically unstable, red for quantum instability, and light/dark green for stability up to the GUT/Planck scales.

We now analyze the phenomenological properties of the points that satisfy the Planck relic density constraint, with the results detailed in Figures~\ref{fig:Z613Graph1} and~\ref{fig:Z613Graph2}. Figure~\ref{fig:Z613Graph1} reveals that the lighter component, $S_A$, is consistently the primary contributor to the total relic density. This dominance appears to be a robust feature of the model, largely insensitive to the value of the coupling $\lambda'_{AB}$. This dynamic has a direct impact on the model's mass spectrum, shown in Figure~\ref{fig:Z613Graph2}. While the points in the viable Higgs resonance region exhibit a healthy, non-degenerate mass splitting, a strong degeneracy ($M_{S_B} \approx M_{S_A}$) emerges for the few points found outside this region. This latter feature significantly limits the potential for distinct experimental signatures, further constraining the model's phenomenological appeal.

The direct detection prospects, shown in Figure~\ref{fig:Z613Graph3A4}, confirm the highly constrained nature of the $\mathbb{Z}_6(13)$ model. The rescaled cross-sections for both components are plotted against their mass, revealing that the phenomenologically viable points are almost exclusively confined to the Higgs resonance region. Once again, the power of the latest LZ (2024) data~\cite{LZ2024} is evident, eliminating vast regions of parameter space that were consistent with previous limits~\cite{Lux-Zeplin2023}. The plot also visually distinguishes between fully excluded points (black) and the `partially excluded' cases (gray), where the combined constraint of Eq.~\eqref{eq:res.2} is decisive. Despite the highly restricted nature of the model, a small cluster of viable points persists above the neutrino floor~\cite{NeutrinoFloor}, suggesting that even this constrained scenario remains testable by future experiments.

\begin{figure}[htbp]
    \centering
    \includegraphics[width=0.45\textwidth]{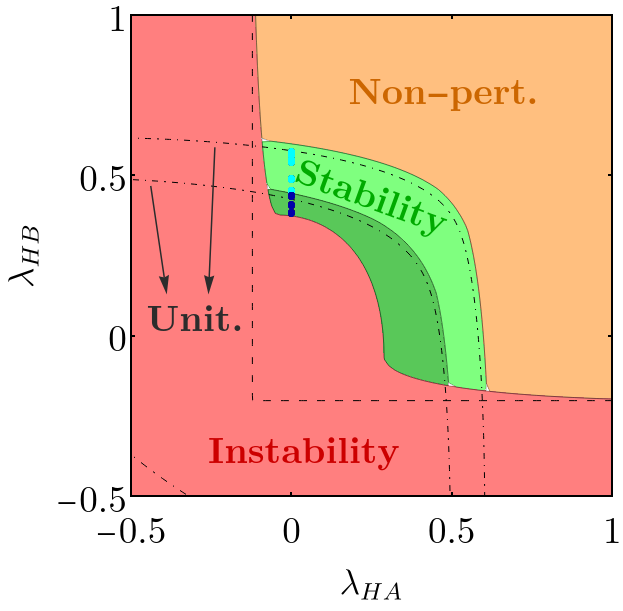}
    \caption{The final, fully constrained parameter space for the $\mathbb{Z}_6(13)$ model, shown in the $(\lambda_{HA}, \lambda_{HB})$ plane for a representative benchmark value of $\lambda'_{AB}$. The plot focuses on the Higgs resonance region, where we find an exceptionally small cluster of points that simultaneously satisfy all theoretical and phenomenological constraints, highlighting the highly restrictive nature of this scenario.}
    \label{fig:Z613Graph5}
\end{figure}

The final synthesis of all constraints for the $\mathbb{Z}_6(13)$ model is presented in Figure~\ref{fig:Z613Graph5}, which maps the phenomenologically viable points onto the fundamental plane of the Higgs portal couplings. Since the theoretical landscape in this model also depends on the value of $\lambda'_{AB}$, this plot is shown for a representative benchmark value of this coupling, focusing on the Higgs resonance region where solutions were found. The result is stark: even within this theoretically consistent slice, the parameter space that simultaneously satisfies all dark matter constraints is exceptionally small. We have verified that this conclusion remains robust, with the viable region staying similarly restricted when other values for $\lambda'_{AB}$ are considered, confirming the highly constrained nature of the model.

\section{Conclusions}
\label{Conc}

In this work, we performed a comprehensive, multi-faceted analysis of three two-component scalar dark matter models stabilized by a $\mathbb{Z}_{2n}$ symmetry: the $\mathbb{Z}_4$, $\mathbb{Z}_6(23)$, and $\mathbb{Z}_6(13)$ scenarios. Our approach was to synthesize the most stringent experimental constraints from cosmology—namely the Planck relic abundance and the latest direct detection limits from LZ—with a rigorous high-scale ``stress test'' of theoretical self-consistency, including vacuum stability and perturbative unitarity, enforced via one-loop Renormalization Group Equations up to the GUT and Planck scales.

Our analysis confirms the $\mathbb{Z}_4$ model as a robust and compelling scenario. Its efficient semi-annihilation channel, driven by a trilinear coupling, opens a wide and viable parameter space for sub-TeV dark matter that evades the tight constraints placed on simpler singlet models. This results in a non-degenerate mass spectrum with promising prospects for discovery in future direct detection experiments.

The $\mathbb{Z}_6(23)$ model presents the most subtle and intriguing result of our study. While an initial analysis reveals a stark tension between low-energy phenomenology and theoretical consistency at the GUT scale, we demonstrated that this is not a fundamental flaw. Instead, this tension is resolved if the model is interpreted as an effective field theory. We found that the entire phenomenologically viable region is theoretically self-consistent, provided the theory possesses a UV completion at or below a cutoff scale of $\Lambda_{\text{max}} \lesssim 10^6$ GeV. This reframes the model from a problematic scenario into one with a concrete, falsifiable prediction: the existence of new physics at the PeV scale, necessary to ensure the ultimate stability of the theory.

Finally, our results confirmed that the $\mathbb{Z}_6(13)$ model is the most constrained of the three. Lacking an efficient semi-annihilation channel, its viable parameter space is severely restricted almost exclusively to the well-known Higgs resonance region ($M_{S_A} \approx M_h/2$), reinforcing its highly fine-tuned nature.

A central conclusion of our work is the indispensable role of a combined analysis. We have shown that high-scale consistency requirements, when confronted with the power of the latest experimental data, act as a crucial theoretical filter. This synergy not only reshapes and constrains the allowed parameter space but can also reveal the true nature of a model, turning apparent inconsistencies into concrete predictions for new physics, as exemplified by our findings for the $\mathbb{Z}_6(23)$ model.

Future work will proceed in two main directions. First, a detailed study of the electroweak phase transition within these models to explore their potential for generating a detectable gravitational wave signal. Second, an investigation into supersymmetric extensions of these multi-component scenarios. This work establishes these $\mathbb{Z}_{2n}$ frameworks as compelling paradigms for exploring the rich interplay between dark matter phenomenology and fundamental theoretical principles.

\appendix

\section{One-Loop Renormalization Group Equations}
\label{RGEs}

This appendix presents the one-loop Renormalization Group Equations (RGEs) used to evolve the model parameters from the top quark mass scale, $M_t = 172.69\ \mbox{GeV}$ \cite{PDG2022}, up to the Planck scale. We define the evolution parameter as $t = \ln \mu$, where $\mu$ is the renormalization scale.

\subsection{Standard Model and Portal Coupling RGEs}

The RGEs for the SM gauge couplings ($g_i$) and the top Yukawa coupling ($y_t$) are unaffected by the new physics at one-loop level:
\begin{align}
(4\pi)^2\frac{dg_i}{dt} &= b_ig_i^3, \quad \text{with} \quad b_i=\left(\frac{41}{10},-\frac{19}{6},-7\right),\label{RGEsgs}\\
(4\pi)^2\frac{dy_t}{dt} &= y_t\left(-\frac{17}{20}g_1^2-\frac{9}{4}g_2^2-8g_3^2+\frac{9}{2}y_t^2\right).\label{RGEyt}
\end{align}
Here, $g_1$, $g_2$, and $g_3$ are the gauge couplings of $\mathrm{U}(1)_Y$, $\mathrm{SU}(2)_L$, and $\mathrm{SU}(3)_C$, respectively, with the GUT-inspired normalization $g_1 \equiv \sqrt{5/3} \, g_Y$.

The RGE for the Higgs self-coupling, $\lambda_H$, receives new contributions from the portal couplings that connect the visible and dark sectors:
\begin{align}
\begin{split}
(4\pi)^2\frac{d\lambda_H}{dt} &= \left( \text{SM terms} \right) + \lambda_{HA}^2 + \frac{1}{2}\lambda_{HB}^2, \\
\text{where} \quad \left( \text{SM terms} \right) &= \frac{27}{200}g_1^4 +\frac{9}{20}g_1^2g_2^2+ \frac{9}{8}g_2^4 -9\lambda_H\left(\frac{1}{5}g_1^2+g_2^2\right) + 24\lambda_H^2 + 12\lambda_H y_t^2 - 6y_t^4.
\end{split}
\end{align}
The RGEs for the portal couplings themselves are common to all three models:
\begin{align}
(4\pi)^2\frac{d\lambda_{HA}}{dt} &= -\frac{9}{2}\lambda_{HA}\left(\frac{1}{5}g_1^2+g_2^2\right) +2\lambda_{HA}\left(6\lambda_H+4\lambda_A+2\lambda_{HA}+3y_t^2\right)+2\lambda_{AB}\lambda_{HB},\\
(4\pi)^2\frac{d\lambda_{HB}}{dt} &= -\frac{9}{2}\lambda_{HB}\left(\frac{1}{5}g_1^2+g_2^2\right) +6\lambda_{HB}\left(2\lambda_H+4\lambda_B+y_t^2\right)+4\lambda_{AB}\lambda_{HA}+4\lambda_{HB}^2.
\end{align}

\subsection{Dark Sector Self-Coupling RGEs}

The RGE for the self-coupling of the real scalar, $\lambda_B$, is the same for all models:
\begin{equation}
(4\pi)^2\frac{d\lambda_{B}}{dt} = 72\lambda_B^2+\lambda_{AB}^2  +\frac{1}{2}\lambda_{HB}^2.
\end{equation}
The RGEs for the remaining dark sector couplings depend on the specific model, reflecting their unique interaction terms. For the self-coupling of the complex scalar, $\lambda_A$, we have:
\begin{equation}
(4\pi)^2\frac{d\lambda_{A}}{dt} = 2(10\lambda_{A}^2 + \lambda_{AB}^2 + \lambda_{HA}^2) + \Delta_A,
\end{equation}
where $\Delta_A = 36\lambda_{S4}^2$ for the $\mathbb{Z}_4$ model, $\Delta_A = 2\lambda_{AB}'^2$ for the $\mathbb{Z}_6(13)$ model, and $\Delta_A = 0$ for the $\mathbb{Z}_6(23)$ model.

Similarly, the RGE for the mixed quartic coupling, $\lambda_{AB}$, is given by:
\begin{equation}
(4\pi)^2\frac{d\lambda_{AB}}{dt} = 8\lambda_{AB}(\lambda_A+3\lambda_B+\lambda_{AB})+ 2\lambda_{HA}\lambda_{HB} + \Delta_{AB},
\end{equation}
where $\Delta_{AB} = 4\lambda_{AB}'^2$ for the $\mathbb{Z}_6(13)$ model, and $\Delta_{AB} = 0$ for the $\mathbb{Z}_4$ and $\mathbb{Z}_6(23)$ models.

Finally, the RGEs for the couplings unique to specific models are:
\begin{align}
(4\pi)^2\frac{d\lambda_{S4}}{dt} &= 24\lambda_{S4}\lambda_A \quad (\text{for } \mathbb{Z}_4 \text{ model}), \\
(4\pi)^2\frac{d\lambda_{AB}'}{dt} &= 12\lambda_{AB}'(\lambda_A+\lambda_{AB}) \quad (\text{for } \mathbb{Z}_6(13) \text{ model}).
\end{align}

\subsection{Experimentally Fixed Initial Conditions}
\label{sec:fixed_params}

The SM gauge couplings, the top Yukawa coupling ($y_t$), and the Higgs quartic coupling ($\lambda_H$) are fixed to their experimentally determined values at the top quark mass scale ($M_t$) \cite{ButtazzoStrumia}. We use the following initial conditions:
\begin{equation}
\label{eq:SM_initial_conditions}
\begin{split}
    & g_1(M_t)=\sqrt{\frac{5}{3}}\times 0.35767, \quad g_2(M_t)=0.64815, \quad g_3(M_t)=1.16499,\\
    & \hspace*{0.15\textwidth} y_t(M_t)=0.93378, \quad \lambda_H(M_t)=0.12628,
\end{split}
\end{equation}
where these values are derived from the following experimental inputs \cite{PDG2022}: $M_h = 125.25 \pm 0.17 \ \text{GeV}$, $M_t = 172.69 \pm 0.30 \ \text{GeV}$, $M_W = 80.377 \pm 0.012 \ \text{GeV}$, $M_Z = 91.1876 \pm 0.0021 \ \text{GeV}$, and $\alpha_s(M_Z) = 0.1180 \pm 0.0009$.

\section{Relic Density}
\label{reldens}

\begin{figure}[htbp]
    \centering

    \begin{subfigure}{0.8\textwidth}
        \centering
        \includegraphics[width=1.0\textwidth]{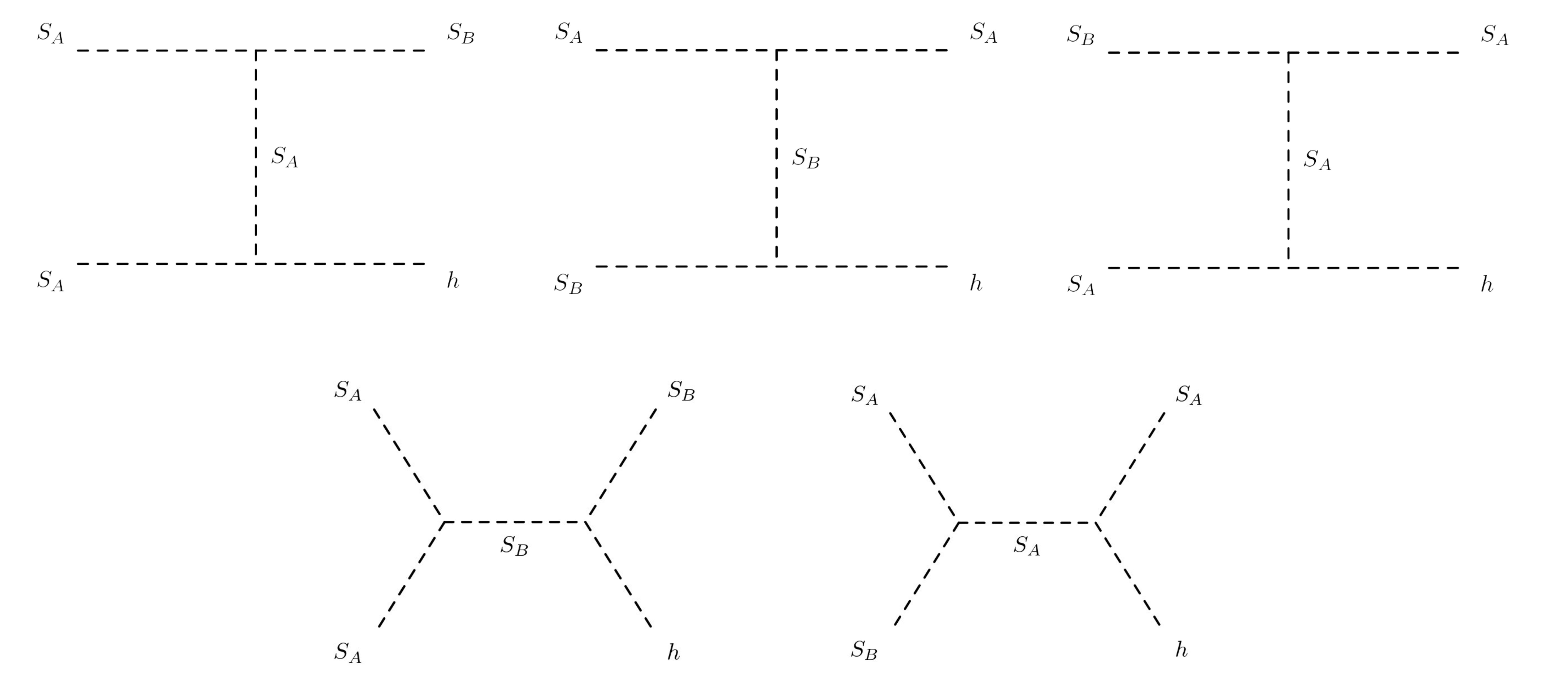}
        \caption{Semi-annihilation processes unique to the $\mathbb{Z}_{4}$ model, driven by the trilinear interaction $\mu_{S1}$.}
        \label{fig:diagramZ4}
    \end{subfigure}
    
    \medskip

    \begin{subfigure}{0.8\textwidth}
        \centering
        \includegraphics[width=0.9\textwidth]{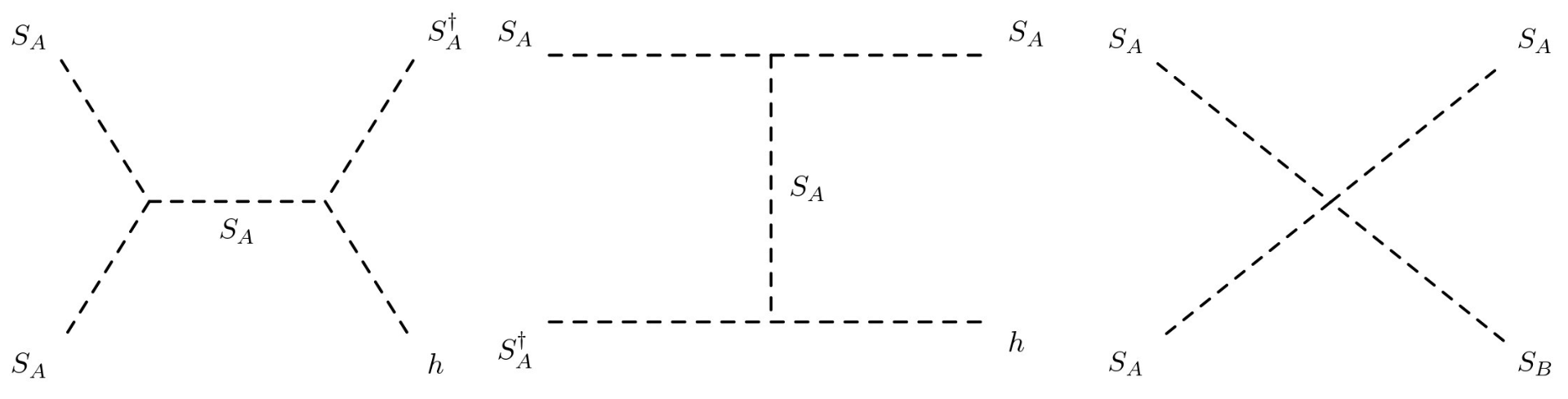}
        \caption{Model-specific interactions for the $\mathbb{Z}_{6}$ scenarios. Left \& Center: Semi-annihilation $S_A S_A \to S_A^{\dagger} h$ in the $\mathbb{Z}_{6}(23)$ model, driven by $\mu_{S2}$. Right: Conversion process $S_A S_A \to S_B S_A^\dagger$ in the $\mathbb{Z}_{6}(13)$ model, driven by $\lambda'_{AB}$.}
        \label{fig:diagramZ6}
    \end{subfigure}
    
    \medskip

    \begin{subfigure}{0.8\textwidth}
        \centering
        \includegraphics[width=0.75\textwidth]{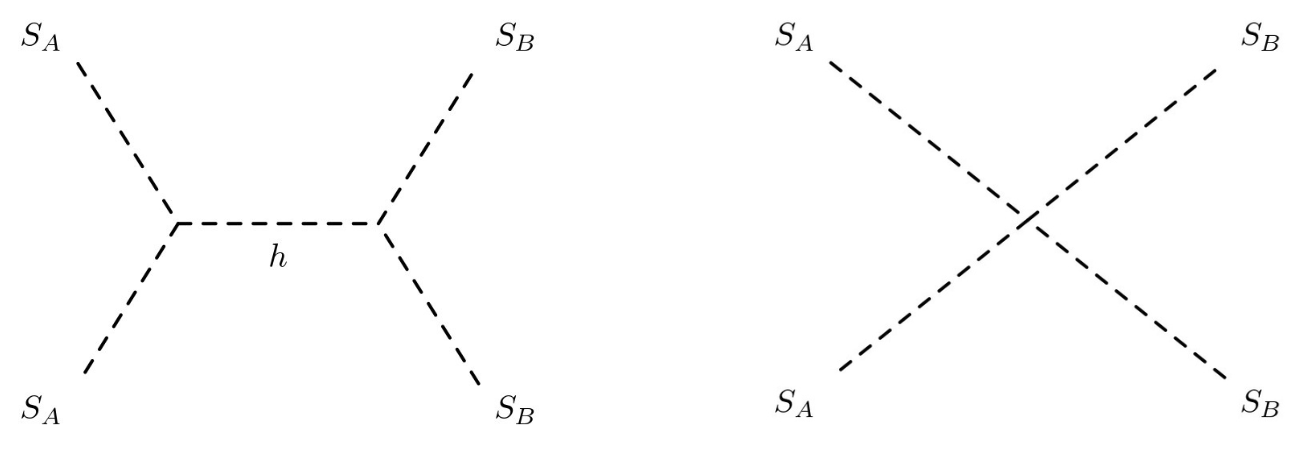}
        \caption{Conversion process $S_A S_A^{\dagger} \leftrightarrow S_B S_B$, common to all models, mediated by the quartic coupling $\lambda_{AB}$.}
        \label{fig:diagramZNconv}
    \end{subfigure}
    
    \medskip

    \begin{subfigure}{0.8\textwidth}
        \centering
        \includegraphics[width=0.9\textwidth]{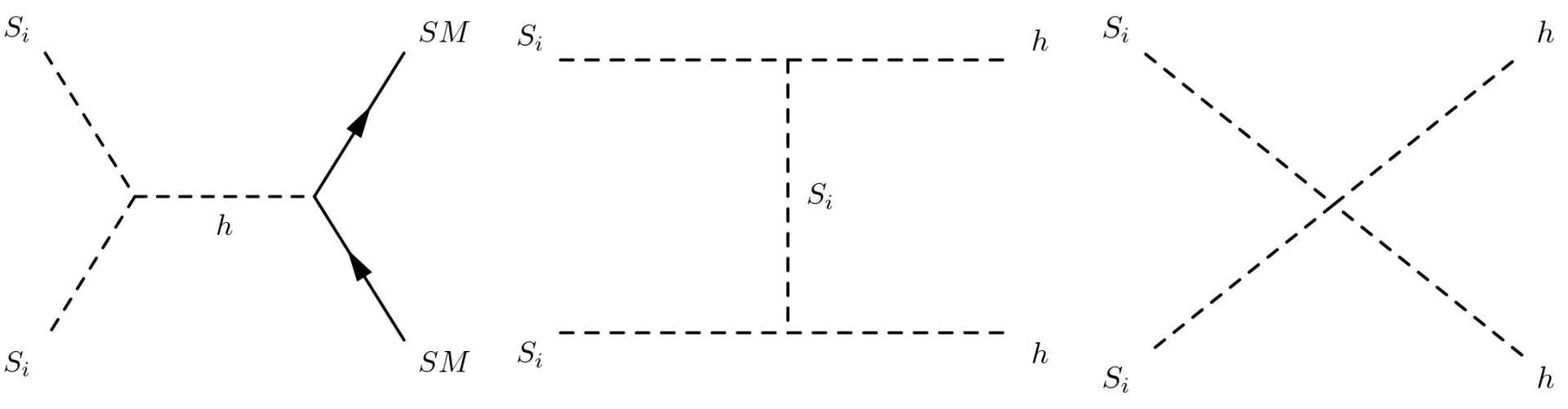}
       \caption{Standard annihilation channels for DM particles $S_i$ ($i = A, B$) into SM final states ($\text{SM}=f, h, W, Z$). These processes, common to all models, are mediated by the Higgs portal couplings $\lambda_{Hi}$.}
        \label{fig:deteccdire}
    \end{subfigure}

    \caption{Feynman diagrams for the key dark matter interaction processes that determine the relic abundance in the $\mathbb{Z}_{2n}$ models. Panels distinguish between processes unique to specific models and those common to all.}
    \label{fig:diagramZ_all}
\end{figure}

This appendix provides the detailed coupled Boltzmann equations used to calculate the relic densities of the DM candidates for each of the $\mathbb{Z}_{2n}$ models discussed in the main text. These equations, which account for all relevant annihilation, co-annihilation, conversion, and semi-annihilation processes shown in Fig.~\ref{fig:diagramZ_all}, form the basis of the numerical analysis performed with \texttt{micrOMEGAs}.

\subsection{\texorpdfstring{$\mathbb{Z}_{4}$}{Z4} Model}
\label{RelicDensZ4Model}

The $\mathbb{Z}_4$-symmetric Lagrangian includes the trilinear term $\frac{1}{2}\mu_{S1} S_{A}^{2} S_{B}$. This interaction introduces a crucial semi-annihilation channel ($S_A S_A \to S_B h$) and imposes a mass hierarchy, $M_{S_B} < 2M_{S_A}$, to ensure the stability of both DM components by forbidding the decay $S_B \rightarrow S_A S_A$. The evolution of the relic densities is thus governed by a combination of standard annihilation into SM particles, Fig.~\ref{fig:deteccdire}, conversion between the dark matter species, Fig.~\ref{fig:diagramZNconv}, and the crucial semi-annihilation channel unique to this model, Fig.~\ref{fig:diagramZ4}. The resulting coupled Boltzmann equations are:
\begin{align}
\dot{n}_{S_A} + 3Hn_{S_A} = &- \langle \sigma v \rangle_{S_A S_A^{\dagger} \to \text{SM SM}} \left(n_{S_A}^2 - n_{S_A}^{\text{eq}\,2} \right)
 - \langle \sigma v \rangle_{S_A S_A^{\dagger} \to S_B S_B} \left(n_{S_A}^2 - n_{S_B}^2 \frac{n_{S_A}^{\text{eq}\,2}}{n_{S_B}^{\text{eq}\,2}} \right) \nonumber \\
 &- \langle \sigma v \rangle_{S_A S_A \to S_B h} \left(n_{S_A}^2 - n_{S_A}^{\text{eq}\,2} \frac{n_{S_B}}{n_{S_B}^{\text{eq}}} \right),
\end{align}
\begin{align}
\dot{n}_{S_B} + 3Hn_{S_B} = &- \langle \sigma v \rangle_{S_B S_B \to \text{SM SM}} \left(n_{S_B}^2 - n_{S_B}^{\text{eq}\,2} \right)
 - \langle \sigma v \rangle_{S_B S_B \to S_A S_A^{\dagger}} \left(n_{S_B}^2 - n_{S_A}^2 \frac{n_{S_B}^{\text{eq}\,2}}{n_{S_A}^{\text{eq}\,2}} \right) \nonumber\\
 &- \langle \sigma v \rangle_{S_B S_A \to S_A^{\dagger} h} \left(n_{S_B}n_{S_A} - n_{S_A}^{\text{eq}}n_{S_B}^{\text{eq}} \right) \label{eq:4.2.1.2} \\
 &+\frac{1}{2} \langle \sigma v \rangle_{S_A S_A \to S_B h} \left(n_{S_A}^2 - n_{S_A}^{\text{eq}\,2} \frac{n_{S_B}}{n_{S_B}^{\text{eq}}} \right).\nonumber
\end{align}

\subsection{\texorpdfstring{$\mathbb{Z}_{6}(23)$}{Z6(23)} Model}
\label{RelicDensZ6(23)Model}

The $\mathbb{Z}_6(23)$ model introduces the cubic self-interaction term $\frac{1}{3}\mu_{S2} S_A^3$. The stability of both $S_A$ and $S_B$ is ensured as two-body decays are forbidden by their charge assignments. Besides standard annihilation processes, the relic density evolution is primarily governed by two additional mechanisms: dark sector conversion and semi-annihilation. Conversion, mediated by the quartic coupling $\lambda_{AB}$, drives number density exchanges of the form $S_A S_A^{\dagger} \leftrightarrow S_B S_B$. Semi-annihilation, enabled by the cubic self-interaction $\mu_{S2}$ in conjunction with the Higgs portal coupling, facilitates processes like $S_A S_A \to S_A^{\dagger} h$.

These mechanisms jointly govern freeze-out dynamics, as shown in Figs.~\ref{fig:diagramZ6}, ~\ref{fig:diagramZNconv} and~\ref{fig:deteccdire}. The coupled Boltzmann system becomes:
\begin{align}
\dot{n}_{S_A} + 3Hn_{S_A} &= -\langle \sigma v \rangle_{S_A S_A^\dagger \to \text{SM SM}} (n_{S_A}^2 - n_{S_A}^{\text{eq}\,2}) - \langle \sigma v \rangle_{S_A S_A^\dagger \to S_B S_B} \left(n_{S_A}^2 - n_{S_B}^2 \frac{n_{S_A}^{\text{eq}\,2}}{n_{S_B}^{\text{eq}\,2}} \right) \nonumber \\
&\quad - \langle \sigma v \rangle_{S_A S_A \to S_A^\dagger h} (n_{S_A}^2 - n_{S_A}^{\text{eq}} n_{S_A} ),\label{eq:Z6-23-A}
\end{align}
\begin{equation}
\dot{n}_{S_B} + 3Hn_{S_B} = -\langle \sigma v \rangle_{S_B S_B \to \text{SM SM}} (n_{S_B}^2 - n_{S_B}^{\text{eq}\,2}) - \langle \sigma v \rangle_{S_B S_B \to S_A S_A^\dagger} \left(n_{S_B}^2 - n_{S_A}^2 \frac{n_{S_B}^{\text{eq}\,2}}{n_{S_A}^{\text{eq}\,2}} \right). \label{eq:Z6-23-B}
\end{equation}

\subsection{\texorpdfstring{$\mathbb{Z}_{6}(13)$}{Z6(13)} Model}
\label{RelicDensZ6(13)Model}

The $\mathbb{Z}_6(13)$-symmetric Lagrangian contains the quartic interaction term $\frac{1}{3}\lambda'_{AB} S_A^3 S_B$, which introduces novel conversion processes between $S_A$ and $S_B$ while requiring the mass hierarchy $M_{S_B} < 3M_{S_A}$ to prevent $S_B \to 3S_A$ decays. This ensures both fields remain viable DM components through chemical equilibrium maintenance.

The processes relevant are conversion processes, arising from the quartic interaction terms $\lambda_{AB}$ and $\lambda'_{AB}$ and, the standard annihilation. These interactions govern the number densities of $S_A$ and $S_B$ (Figs.~\ref{fig:diagramZ6}, ~\ref{fig:diagramZNconv} and ~\ref{fig:deteccdire}). These interactions modify the coupled Boltzmann system as:
\begin{align}
\dot{n}_{S_A} + 3Hn_{S_A} &= -\langle \sigma v \rangle_{S_A S_A \to \text{SM SM}} \left(n_{S_A}^2 - n_{S_A}^{\text{eq}\,2} \right)
- \langle \sigma v \rangle_{S_A S_A^{\dagger} \to S_B S_B} \left(n_{S_A}^2 - n_{S_B}^2 \frac{n_{S_A}^{\text{eq}\,2}}{n_{S_B}^{\text{eq}\,2}} \right) \nonumber
\\
&\quad - \langle \sigma v \rangle_{S_A S_A \to S_B S_A^{\dagger}} \left(n_{S_A}^2 - n_{S_A}^{\text{eq}} n_{S_A} \frac{n_{S_B}}{n_{S_B}^{\text{eq}}} \right),
\end{align}
\begin{align}
\dot{n}_{S_B} + 3Hn_{S_B} &= -\langle \sigma v \rangle_{S_B S_B \to \text{SM SM}} \left(n_{S_B}^2 - n_{S_B}^{\text{eq}\,2} \right)
- \langle \sigma v \rangle_{S_B S_B \to S_A S_A^\dagger} \left(n_{S_B}^2 - n_{S_A}^2 \frac{n_{S_B}^{\text{eq}\,2}}{n_{S_A}^{\text{eq}\,2}} \right)
\nonumber \\
&\quad - \frac{1}{2} \langle \sigma v \rangle_{S_B S_A \to S_A^\dagger S_A^\dagger} \left(n_{S_B}n_{S_A} - n_{S_A}^2 \frac{n_{S_B}^{\text{eq}}}{n_{S_A}^{\text{eq}}} \right).
\end{align}

\section*{Acknowledgements}

\noindent
 B. L. Sánchez-Vega, J. P. Carvalho-Corrêa, and I. M. Pereira thank the National Council for Scientific and Technological Development of Brazil, CNPq, for financial support through Grants n$^{\circ}$ 311699/2020-0, 141118/2022-9, and 161469/2021-3, respectively. A. C. D. Viglioni thanks
the Coordination for the Improvement of Higher Educational Personnel, CAPES, for financial support.

\bibliographystyle{spphys}
\bibliography{Bibliography}

\end{document}